%% file: czti_fast_transients.tex
\documentclass{jaa}
\usepackage{natbib}
\usepackage{xcolor}

\definecolor{xlinkcolor}{cmyk}{1,1,0,0}

 \usepackage[bookmarks=false,         % show bookmarks bar?
     pdfnewwindow=true,      % links in new window
     colorlinks=true,    % false: boxed links; true: colored links
     linkcolor=xlinkcolor,     % color of internal links
     citecolor=xlinkcolor,     % color of links to bibliography
     filecolor=xlinkcolor,  % color of file links
     urlcolor=xlinkcolor,      % color of external links
final=true
 ]{hyperref}

\usepackage{graphicx}
\usepackage{lastpage}

\usepackage{adjustbox}
\usepackage{multirow}
\graphicspath{ {./Figures/} }

\usepackage{tabularx}
\usepackage{booktabs}
\usepackage{ltablex}
\usepackage{makecell}

\usepackage{caption}
\usepackage{subcaption}
\newcommand{\degr}{\ensuremath{^\circ}}
\newcommand{\asat}{{\em AstroSat}}
\newcommand{\tnt}{\ensuremath{\mathrm{T}_{90}}}
\newcommand{\sw}[1]{\texttt{#1}}
\newcommand{\edited}[1]{\textit{\color{orange}}}

\begin{document}\sloppy

%%paper title
%%For line breaks \\ can be used within title
\title{The Search for Fast Transients with CZTI }

%%author names are separated by comma (,)
%%use \and before the last author name
%%use a * along with the number separated by comma
%% for the  author for correspondence
%%\textsuperscript{number} is used for affiliation
%%\affilOne, \affilTwo etc., upto \affilTwentyfive is possible
%%Please note the first letter after \affil is capitalised in the command
%%

\author{\href{https://orcid.org/0000-0003-4531-1745}{Y. Sharma}\textsuperscript{1,2}, \href{https://orcid.org/0000-0002-8869-8137}{A. Marathe}\textsuperscript{2,3}, 
\href{https://orcid.org/0000-0002-6112-7609}{V. Bhalerao}\textsuperscript{2,*},
\href{https://orcid.org/0000-0002-6429-8139}{V. Shenoy}\textsuperscript{2}, 
\href{https://orcid.org/0000-0003-3630-9440}{G. Waratkar}\textsuperscript{2}, \href{https://orcid.org/0000-0003-4942-6873}{D. Nadella}\textsuperscript{3}, 
\href{https://orcid.org/0000-0002-8239-9670}{P. Page}\textsuperscript{2}, \href{https://orcid.org/0000-0002-4961-4690}{P. Hebbar}\textsuperscript{2,4}, 
\href{https://orcid.org/0000-0003-1501-6972}{A. Vibhute}\textsuperscript{5}, \href{http://orcid.org/0000-0003-3352-3142}{D. Bhattacharya}\textsuperscript{5},
\href{https://orcid.org/0000-0003-0833-0533}{A. R. Rao}\textsuperscript{6}, \href{https://orcid.org/0000-0002-2050-0913}{S. Vadawale}\textsuperscript{7}}
\affilOne{\textsuperscript{1} Division of Physics, Mathematics, and Astronomy, California Institute of Technology, Pasadena, CA 91125, USA\\}
\affilTwo{\textsuperscript{2} Indian Institute of Technology Bombay, Powai, Mumbai 400076, India.\\}
\affilThree{\textsuperscript{3} National Institute of Technology Karnataka, Surathkal, Mangalore 575025, India.\\}
 \affilFour{\textsuperscript{4} University of Alberta, Edmonton, AB, T6G 2E1, Canada.\\}
 \affilFive{\textsuperscript{5} Inter-University Centre for Astronomy and Astrophysics, P. O. Bag 4, Ganeshkhind, Pune 411007, India.\\}
 \affilSix{\textsuperscript{6} Tata Institute of Fundamental Research, Homi Bhabha Road, Mumbai 400005, India.\\}
 \affilSeven{\textsuperscript{7} Physical Research Laboratory, Ahmedabad 380009, India.\\}

%Yashvi, Aditi, Varun, Vedant, Drishika, Gaurav, Pranav, Pavan, Ajay, Dipankar, Rao, Santosh

%%escape two column mode for title, affiliation and abstract
%%by giving \twocolumn command as shown

\twocolumn[{

\maketitle

%%include \corres to print the corresponding author Email id
\corres{varunb@iitb.ac.in}

%%include \msinfo for
%%manuscript information such as
%%received, revised and accepted dates
%%
% \msinfo{1 January 2015}{1 January 2015}

%%abstract
\begin{abstract}
The Cadmium Zinc Telluride Imager on AstroSat has proven to be a very effective all-sky monitor in the hard X-ray regime, detecting over three hundred GRBs and putting highly competitive upper limits on X-ray emissions from gravitational wave sources and fast radio bursts. We present the algorithms used for searching for such transient sources in CZTI data, and for calculating upper limits in case of non-detections. We introduce CIFT: the CZTI Interface for Fast Transients, a framework used to streamline these processes. We present details of 88 new GRBs detected by this framework that were previously not detected in CZTI.
\end{abstract}

%%insert keywords separated by 3 hyphens using \keywords{words}

}] %%close the twocolumn escape here

%%include \doinum{number}for the DOI number in the header
%%include \volnum{number} for the volume number in the header
%%include \year{yyyy} for  year of publication in the header
%%include \pgrange{num--num} page range of article in the header
%%include \artcitid{num} for the article citation id
%%include \lp to print last page of the article
%%include \setcounter{page}{pagenum} for the exact starting page of the article

\doinum{12.3456/s78910-011-012-3}
\artcitid{\#\#\#\#}
\volnum{000}
\year{0000}
\pgrange{1--}
\setcounter{page}{1}
\lp{\pageref{LastPage}}

\section{Introduction}
%\rough{Astrosat, CZTI, how we can do transients}
The Cadmium Zinc Telluride Imager \citep[CZTI][]{czti} is a high-energy coded aperture mask instrument on board \asat~\citep{astrosat}
\edited{CZTI comprises of four independent, identical quadrants giving a total physical area of 976~$cm^2$. Each quadrant consists of a 4$\times$4 array of 5~mm thick Cadmium Zinc Telluride detectors, giving good sensitivity in the 20-200~keV energy range and an energy resolution of 11\% at 60~keV. In nominal operations, all incident photons are saved in event-mode with 20~$\mu$s resolution.} 
While primary coded field of view of CZTI is 4.6\degr$\times$4.6\degr, the collimators and support structure of CZTI become increasingly transparent to radiation at energies above $\sim$100~keV, making it sensitive to sources all over the sky. \edited{The off-axis sensitivity depends on the effective area, which in turn is a strong function of energy and direction. Details of the effective area calculations are presented in \citet{mate_this_issue}}. As there are very few bright sources in this energy range, the net contribution of off-axis sources is small and simply manifests itself as a slightly elevated background.

% \edited{The 976~$cm^2$ geometric area consists of 4 quadrants with 16 detector modules arranged in 4$\times$4 matrix. Each module has 256 pixelated contacts arranged in 16x16 array and a  digital readout which records the energy of incident photon and its pixel location. With an angular resolution of 17' and an energy resolution of 6.5~keV (11\% at 60~keV), CZTI makes for a good imaging and spectroscopy instrument in hard X-ray. Moreover, the timing resolution of individual CZT modules is $\le1 \mu s$ with negligible dead time. From expected bright source count rates, the timing resolution of CZTI is determined to be 20 $\mu s$. CZTI also has polarimetry capability which is explained in detail in \citet{czti,rao2017cadmium}.}

% \edited{Although the primary objective of CZTI is measuring hard X-ray spectrum of bright X-ray sources, the collimators and support structure of it become increasingly transparent to radiation at energies above $\sim$100~keV, making it sensitive to sources all over the sky. CZTI as a wide-angle monitor covers roughly one-third of the sky at all times. \citep[Figure 5]{czti}. As there are very few bright sources in this energy range, the net contribution of off-axis sources is small and simply manifests itself as a slightly elevated background (for details see \citep{mate_this_issue}).}

A special exception to this are bright, short-duration transient sources like gamma ray bursts (GRBs). GRBs with their high brightness and short durations (seconds to minutes) manifest themselves as an increase in the count rates in CZTI. Starting from the first GRB detection on the day the instrument was powered on \citep[GRB~151006A;][]{bbr+15,Rao2016a} CZTI has detected 325 GRBs in the five years since launch. On the other hand, the lack of a measurable change in count rates corresponding to a transient event can be mapped to an upper limit on the flux of the transient. With this technique, we have obtained stringent upper limits on X-ray emission from Fast Radio Bursts \citep{2020ApJ...888...40A}, as well as from gravitational wave sources \citep{bkb+17}.

In this paper, we describe the methods used for searching for such sources (called fast transients hereafter). In \S\ref{sec:dataprep} we discuss the pre-processing of data for our searches. In \S\ref{sec:triggered} we discuss the search for ``known'' transients, where the time and possibly location are known from other sources. We also discuss methods for putting upper limits on the flux from such transients in case they are not detected in data. In \S\ref{sec:search} we discuss in detail the algorithms, software, and the interface developed for searching for transients in all of CZTI data. In \S\ref{sec:results} we discuss the performance of our software, and present the 88 transients detected in our searches. We conclude by discussing future improvements in \S\ref{sec:conc}.

\section{Preparing the data}\label{sec:dataprep}
%\rough{change bc / ds thresholds, make LC with livetime corrections, detrend}

The CZTI data reduction pipeline\footnote{CZTI pipeline: \mbox{\url{http://astrosat-ssc.iucaa.in/?q=cztiData}}} is designed for imaging and spectroscopy of sources in the primary field of view. There are two particular operations in the pipeline that are detrimental to the search and analysis of fast transients. First, the pipeline discards data from time intervals when the on-axis source being targeted by \asat\ is occulted behind earth --- though CZTI might still detect fast transients that are located elsewhere in the sky. Second, sections of data where the count rates in detectors rise above a certain value are discarded as noisy: thus suppressing bright transients. For fast transient searches, we overcome these issues by changing a few pipeline parameters --- thus ensuring that final data products are still compatible with any post-processing software. We follow the standard procedure to obtain Level-2 ``bunch cleaned'' data created by \texttt{cztbunchclean}. Next, when selecting good time intervals with \texttt{cztgtigen}, we change the config file \texttt{mkfThresholds.txt} to remove the earth occult condition (the ELV parameter), which would have discarded data when the on-axis target was occulted by the earth.The next stage is to reject noisy sections of data using \texttt{cztpixclean}. The default settings of \texttt{cztpixclean} discard intervals where a single pixel has more than 2  counts per second, or where a module has more than 35 counts per second. To ensure that this step does not discard bright transients, we raise the detector count threshold to 1000 and the pixel count threshold to 100. Finally, we run \texttt{cztevtclean} to obtain cleaned event files. Since our processing is done independently for each quadrant, we use the \texttt{\textunderscore quad\textunderscore clean.evt} files.

The next stage is to create light curves for each quadrant. Here we have to carefully correct for various sources of dead time in the instrument: for instance quadrant-level dead time (0.3~s dead time for collecting housekeeping data every 100~s), and module-wise dead time (arising from discarding particle--induced photon bunches). We use the pipeline module \texttt{cztbindata} to consider all these factors to correctly calculate the dead time for each time bin used. For certain searches, we also limit select the photon energy ranges in this step.

The final step in data preparation is to remove the orbit-induced trends in the background. As \asat\ is in low earth orbit, the satellite sees a variable background count rate over different parts of the earth, rising near the South Atlantic Anomaly (SAA). We see that the background variations are relatively smooth, over timescales of hundreds of seconds. But, if a transient event were to evolve on comparable or longer timescales, we would not be able to distinguish it from background variations. Fortuitously, most transients of interest have timescales of tens of seconds or shorter. Hence, we can fit a smooth trend to the data and subtract it, effectively making the data ``background-free'' and greatly simplifying the task of transient detection. We have tested two methods for de-trending the data: in the first method, the trend is estimated by using a running median filter of 100 second width. In the second method, we estimate the background using a second order Savitzky-Golay (savgol) filter of 100 second width \citep[for details see][]{2020ApJ...888...40A}. Both trend estimates work well, and hence both are coded into our software. In preliminary testing, the savgol filter yielded better results for transient searches, hence it is set as the default filter.
%\todo{Check order of savgol filter: 2 or 3?}\rough{Aditi: The default filter order is 2, can be changed}
% Gaurav: this systematic search will be one of your tasks. What should be the type, order and width of the filter for best de-trending?

%\todo{Use the word transient everywhere, not grb}

\section{Triggered searches}\label{sec:triggered}
In CZTI data analysis, searches for fast transients are broadly categorised into two types: ``Triggered'' and ``blind''. Triggered searches are cases where the time of a transient, and possibly its position, are already known. For such cases, a qualitative search is carried out by pre-processing the data followed by visual examination. Blind searches, that are more quantitative, are discussed in \S\ref{sec:search}.

\subsection{Method}\label{sec:visual}
\begin{figure*}[!t]
\centering
\begin{subfigure}{0.32\linewidth}
\centering
	\includegraphics[width=\columnwidth]{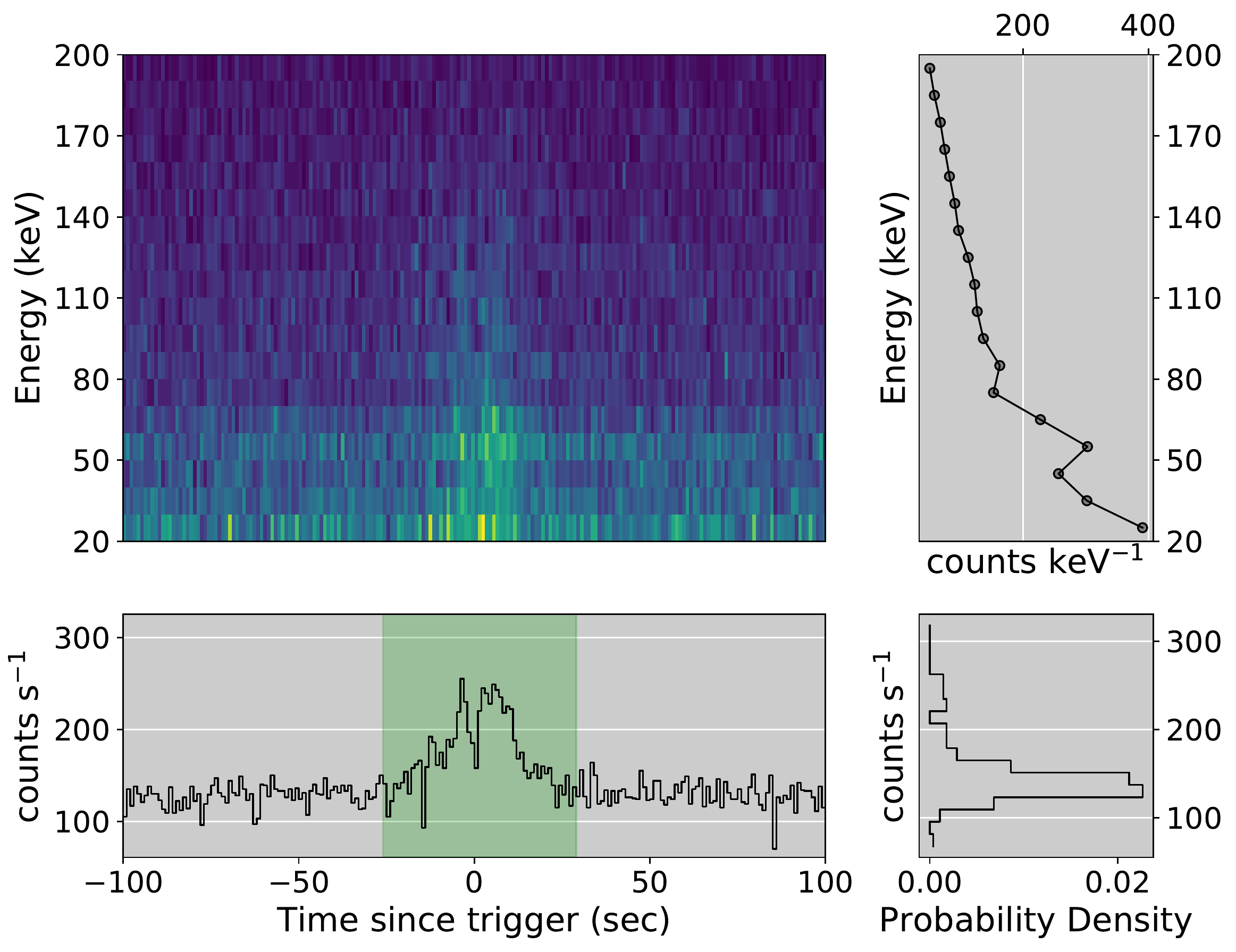}
	\caption{Raw spectrogram}\label{fig:spec_raw}
\end{subfigure} \hfill
\begin{subfigure}{0.32\linewidth}
\centering
	\includegraphics[width=\columnwidth]{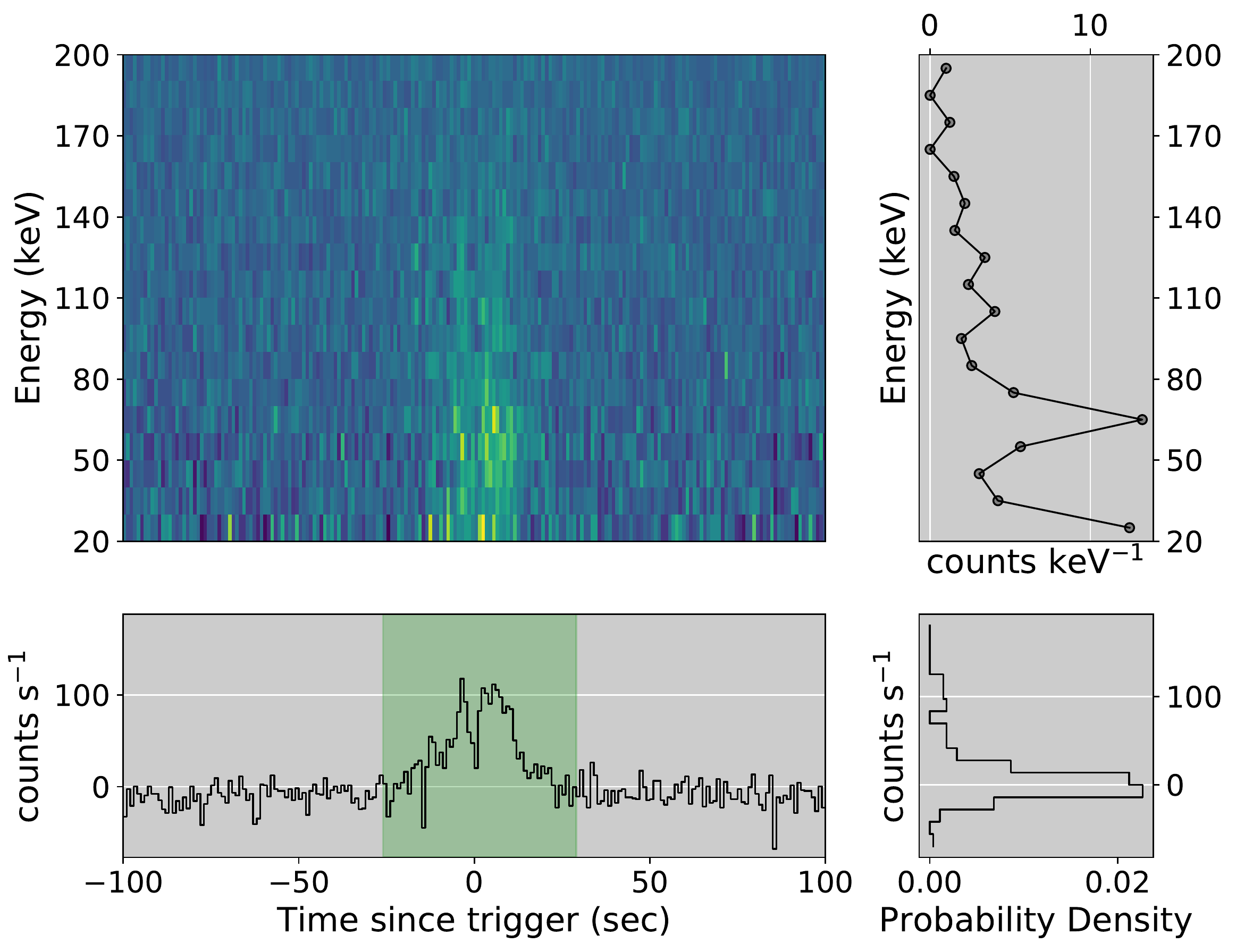}
	\caption{Mean-subtracted spectrogram}\label{fig:spec_sub}
\end{subfigure} \hfill
\begin{subfigure}{0.32\linewidth}
\centering
	\includegraphics[width=\columnwidth]{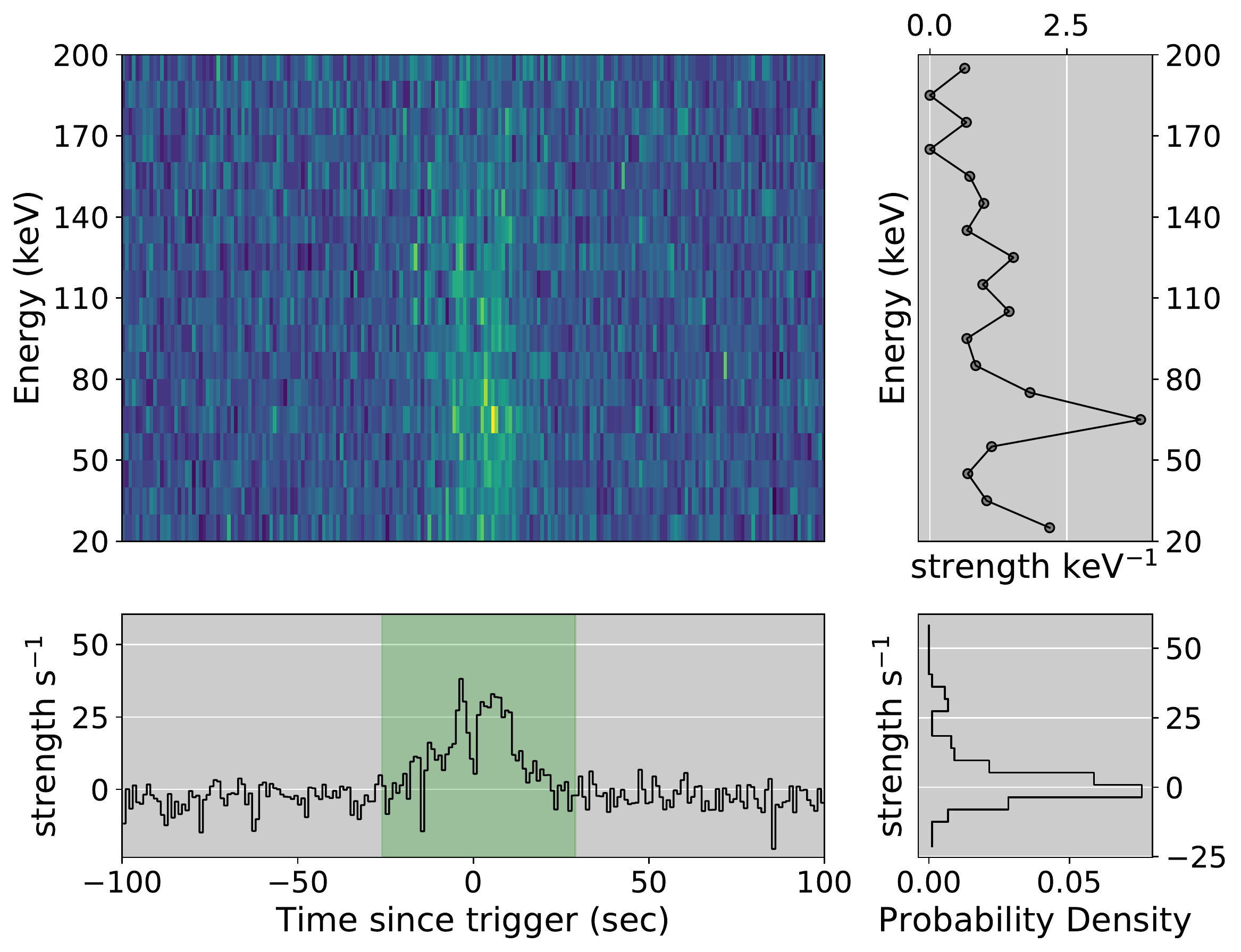}
	\caption{Normalised spectrogram} \label{fig:spec_norm}
\end{subfigure}
\caption{Spectrograms for Quadrant C data for GRB 200306A, utilised in visual inspection of transient candidates (\S\ref{sec:visual}). Panel \subref{fig:spec_raw}: the upper left frame shows raw data, binned in 1~s and 10~keV bins along the X and Y axes respectively. The upper right frame shows the spectrum, obtained by summing the spectrogram along the X axis. The lower left frame shows the light curve, obtained by summing the spectrogram along the Y axis. The lower right frame shows the distribution of count rates in the light curve. Panel \subref{fig:spec_sub}: mean-subtracted spectrogram, obtained by subtracting the average spectrum from each time bin. The four frames are analogous to panel \subref{fig:spec_raw}. Panel \subref{fig:spec_norm}: mean subtracted and sigma-normalised spectrogram. Note that the transient is brightest at the lowest energy bins (Panel \subref{fig:spec_raw}), but since those energies also have a higher sigma, the transient is statistically most significant around 60~keV (Panel \subref{fig:spec_norm}).}
\label{fig:spectrogram}
\end{figure*}

Triggered searches start with pre-processing the data as discussed in \S\ref{sec:dataprep}, up to the creation of cleaned event files. We then create ``spectrograms'' or ``time-energy plots'': two dimensional histograms of the event data, and visually examine them for the transient (Figure~\ref{fig:spectrogram}a). By default, the energy axis is binned in 10~keV bins from 20--200~keV. Searches are carried out by binning the time axis in 0.1~s, 1~s, and 10~s bins. We also calculate two further variants of this spectrogram to aid visual searches: we calculate the mean spectrum and subtract it from each time bin, thus highlighting any transient variations (Figure~\ref{fig:spectrogram}b). In the third step, we take these mean-subtracted spectrograms and normalise the light curve in each energy bin by its standard deviation (Figure~\ref{fig:spectrogram}c). This de-weights noisy energy bands, and gives a rough idea of the statistical significance of any transient.

Light curves from a single quadrant occasionally show noise spikes which look similar to astrophysical transients. These events --- often caused by charged particles or electronic noise --- typically occur at low energies ($\lesssim 50$~keV). Since the four quadrants of CZTI are electronically independent, the electronic noise events are always caused in just a single quadrant. Such noise candidates are readily rejected by requiring that any transient is considered ``detected'' only if it is detected across multiple energy bins, and seen in more than one of the four independent quadrants of CZTI. Track-like events created by charged particles can sometimes be simultaneously seen in multiple quadrants. Such cases are always of short duration ($<1$~s), and can be discarded based on their track-like count distributions in the detector plane. Overall, four quadrant detections of transients are most unambiguous, but detections coincident in three or two quadrants are also considered acceptable if they pass the above cuts, are bright and broadband.

\begin{figure*}[htp]
\centering
\begin{subfigure}{0.49\linewidth}
\centering
	\includegraphics[width=\columnwidth]{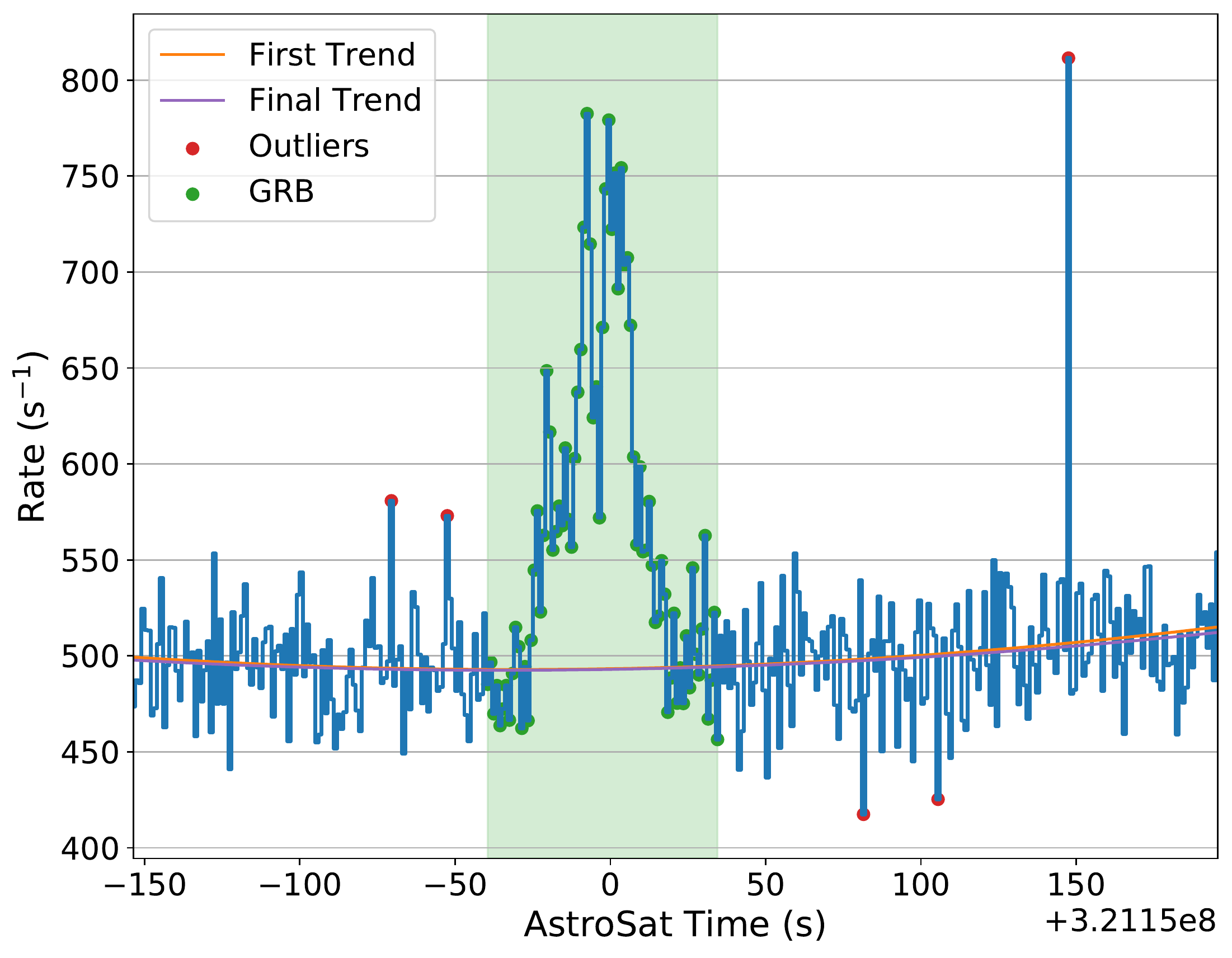}
	\caption{Raw light curve}\label{fig:t90raw}
\end{subfigure}\hfill
\begin{subfigure}{0.49\linewidth}
\centering
	\includegraphics[width=\columnwidth]{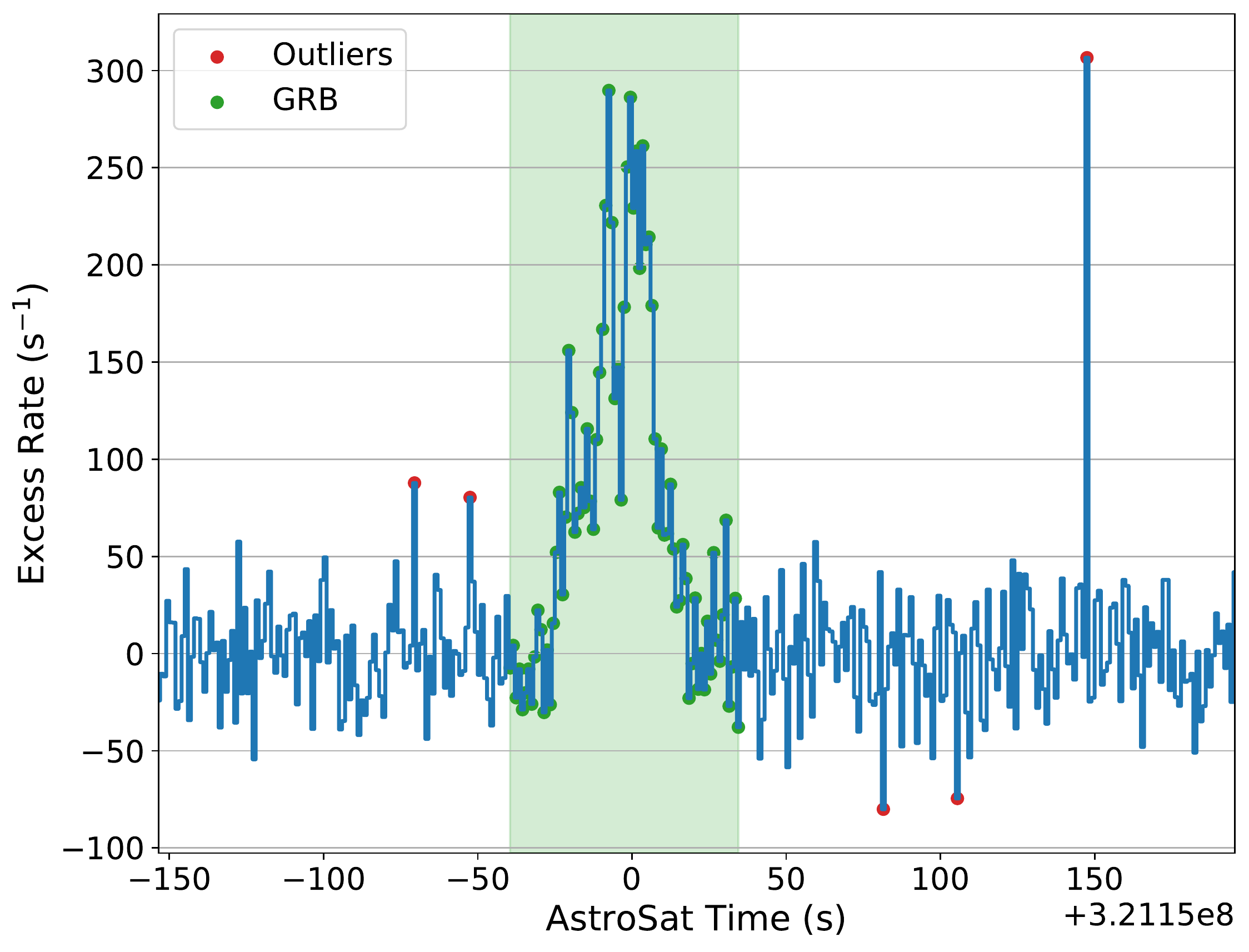}
	\caption{De-trended light curve}\label{fig:t90detrended}
\end{subfigure}

\begin{subfigure}{0.49\linewidth}
\centering
	\includegraphics[width=\columnwidth]{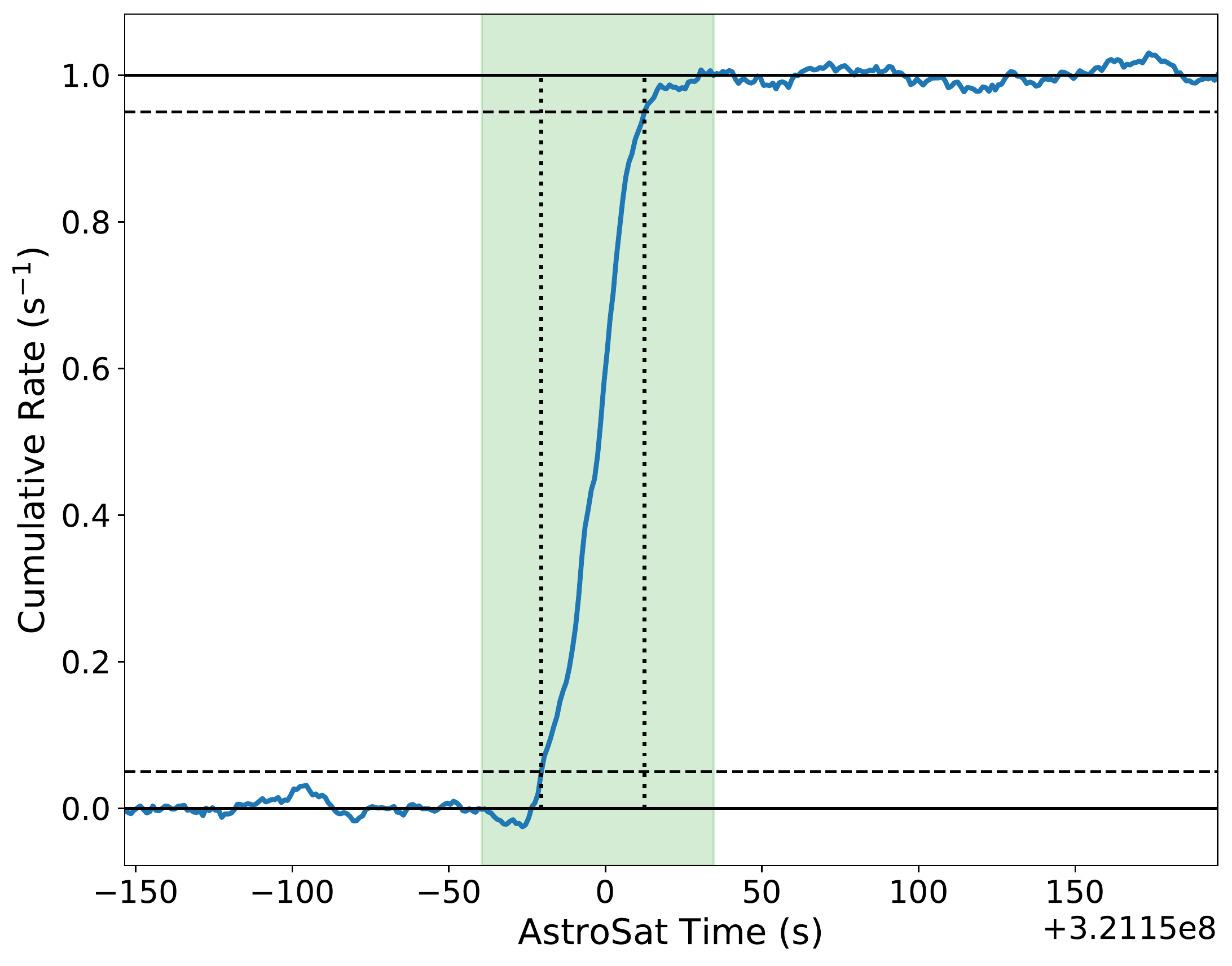}
	\caption{Normalised cumulative count rate}\label{fig:t90cum}
\end{subfigure}\hfill
\begin{subfigure}{0.49\linewidth}
\centering
	\includegraphics[width=\columnwidth]{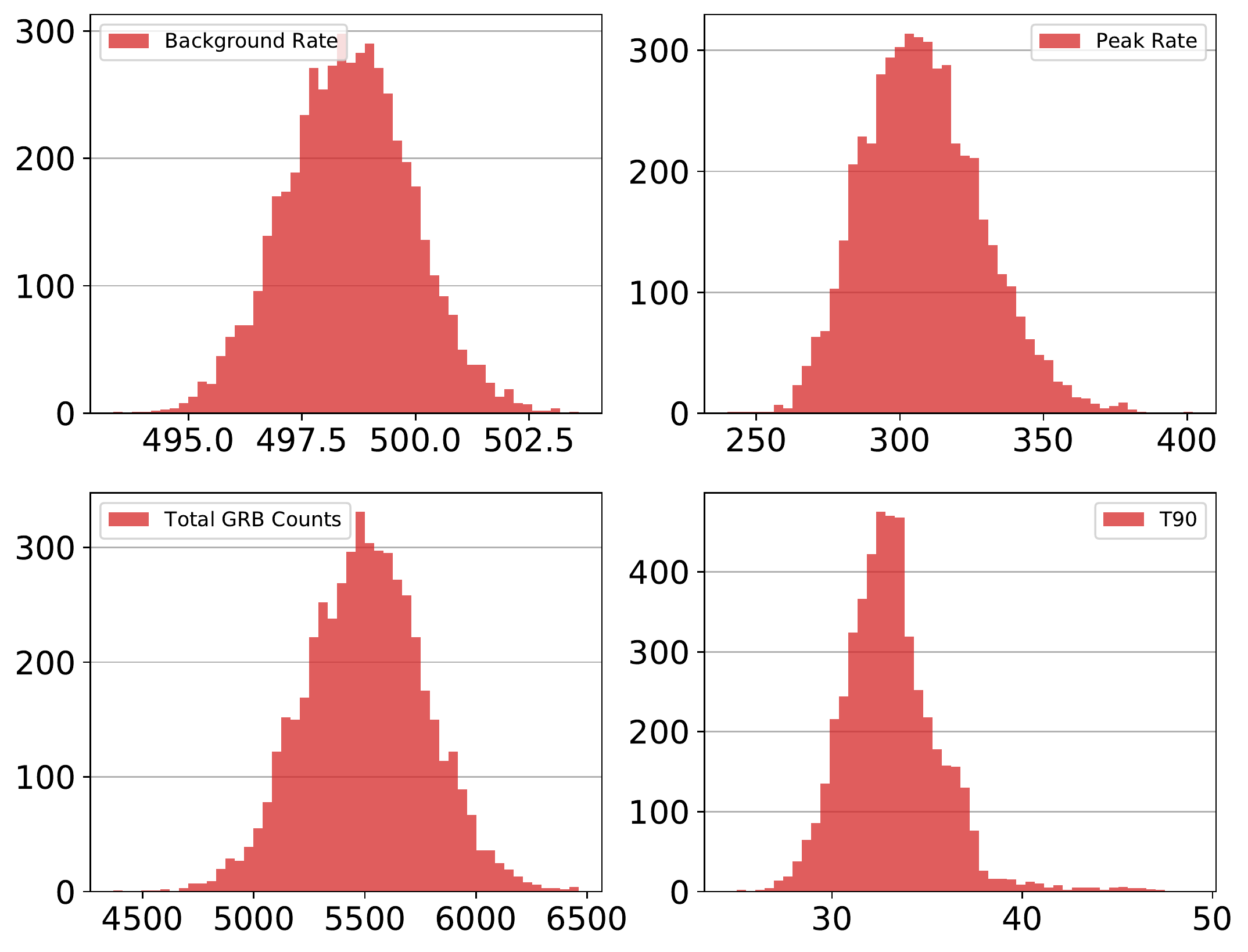}
	\caption{Distributions of calculated parameters}\label{fig:t90dist}
\end{subfigure}
\caption{Calculation of transient properties, illustrated with the light curve of GRB~200306A. Panel \subref{fig:t90raw}: The raw 20--200~keV light curve summed across four quadrants, with the transient region marked in green. An initial background trend (orange) is fit to the background outside the transient region, and refined (purple) with sigma-clipping outlier rejection. Outliers are marked with red circles. The mean value of the refined trend is reported as the background count rate, $R_b$. Panel \subref{fig:t90detrended}: De-trended light curve obtained by subtracting the background trend. The peak count rate ($R_p$) and total counts ($C_\mathrm{tot}$) are measured from this de-trended light curve. Panel \subref{fig:t90cum}: A cumulative light curve calculated from \subref{fig:t90detrended}, normalised such that the median pre-- and post--transient values are 0 and 1 respectively. Dashed lines indicate the points where data cross the 5\% and 95\% levels, which is used to calculate \tnt. Panel \subref{fig:t90dist}: multiple light curves are generated from \subref{fig:t90raw} by assuming Poisson noise distribution, and the four parameters are measured for each of these. The four frames, clockwise from upper left, show distributions of $R_b$, $R_p$, \tnt, and $C_\mathrm{tot}$ obtained from these light curves. These distributions are used to define 90\%\ confidence error bars for the parameters actually measured in \subref{fig:t90detrended} and \subref{fig:t90cum}. For GRB~200306A, we get $R_b = 495_{-3}^{+4}$~counts~s$^{-1}$, $R_p = 289_{-19}^{+51}$~counts~s$^{-1}$, $\tnt = 32^{+4}_{-7}$~s, and $C_\mathrm{tot} = 5444_{-1023}^{+449}$~counts}
\label{fig:t90_1}
\end{figure*}

CZTI also has Caesium Iodide scintillators as anti-coincidence ``veto'' detectors, to reject particle events. Veto detector spectra are sampled once per second, and downlinked along with CZT data. We generate similar spectrograms and light curves for veto data and repeat the transient search. Since data are intrinsically binned at 1~s, the default searches are carried out only at 1~s and 10~s timescales.

These searches are typically run by the Payload Operations Centre (POC) at IUCAA. Transients detected thus are reported in GCN circulars \citep[for instance][etc]{2020GCN.28451....1G,bkb+16a} and announced on the CZTI GRB page at \url{http://astrosat.iucaa.in/czti/?q=grb}, along with the associated spectrograms.

\subsection{Transient properties}\label{sec:properties}
For every detected transient, we estimate its duration (\tnt), peak rate ($R_p$) above background ($R_b$), and the total counts ($C_\mathrm{tot}$). We create a  combined 20--200~keV light curve from all quadrants that show a clear detection of the transient. ``Pre-transient'' and ``Post transient'' sections of the light curve are visually identified, and the background is estimated by fitting a quadratic to these. The best-fit quadratic is subtracted from the data to obtain a background-free light curve, and counts are summed to create a cumulative light curve. The post-transient part of this curve gives a measure of the total counts in the transient. The time taken for the cumulative curve to rise from 5\% to 95\% of the total counts is the \tnt\ duration of the transient (Figure~\ref{fig:t90_1}). These details are included in the published GCN circulars.

\edited{We use a Monte-Carlo approach to estimate the uncertainties in the GRB parameters. We assume that observed photons follow a Poisson distribution, and for simplicity use the observed number of photons in each bin as the rate ($\lambda$) parameter for the Poisson distribution in that bin. We create 5000 simulated light curves by drawing photons from such Poisson distributions for each bin, and measure the four parameters $T_{90}$, $R_p$, $R_b$ and $C_\mathrm{tot}$ for each simulated light curve. We use 5\%--95\% range in the histograms of these parameters (Figure~\ref{fig:t90dist}) as the 90\% credible intervals. For instance, the observed light curve of GRB~200306A yields $T_{90} = 32$~s (Figure~\ref{fig:t90cum}), while the central 90\% credible region is from 25~s to 36~s (Figure~\ref{fig:t90dist}). We report this as $T_{90} = 32_{-7}^{+4}$~s.}

% \edited{Our method to calculate $T_{90}$, $R_p$, $R_b$ and $C_\mathrm{tot}$ is based on a first principle's approach which assumes that the observed photons are drawn from a Poisson distribution of the unknown true rate in each bin. In order to quantify the uncertainty in these parameters, we proceed as follows. We assume that the observed lightcurve is obtained by having a Poisson realisation of a true unknown count rate in every count bin. As a first approximation, we assume that the observed lightcurve is indeed the true underlying lightcurve. We then create 5000 new lightcurves by using Poisson distributions with the mean values given by our lightcurve. For each parameter, the standard deviation measured for these 5000 lightcurves are reported as error bars (Figure~\ref{fig:t90dist}). An important point to note is that the reported value of $T_{90}$ is the $T_{90}$ measured for the actual lightcurve, with the interval given by the 5\% to 95\% of the cumulative lightcurve.}

%To estimate the uncertainty in (T$_{90}$), we generate 5000 new instances of the combined light curve, by assuming that the data follow a Poisson distribution with the rate parameter equal to observed counts. The process is repeated for each of the simulated light curves, and standard deviation of all the $T_{90}$ duration of parameters are reported as error bars (Figure~\ref{fig:t90dist}). 

%\todo{Keep figures ready showing how this is done for a *good* GRB, let's see if we will actually include them.}

    \subsection{Count rate limits for non-detections}\label{sec:nondet}

In cases where no transient is seen, we can place upper limits on the maximum counts received from the transient that would be consistent with noise. Since the mean background level varies through the orbit, we cannot use a direct rate. Instead, we de-trend the data as discussed in \S\ref{sec:dataprep}. %, and use de-trended count rates.
In addition, due to the noise spikes discussed in \S\ref{sec:visual}, the distribution of count rates deviates significantly from a simple Poisson or Normal distribution. In particular, there is a large tail of positive counts with respect to the mean rate which can mimic transient signals. To overcome this hurdle of an unmodeled count rate distribution, we estimate the upper limits (hereafter referred to as cutoff rates) using data from nearby orbits. The method is based on the assumption that the rate of astrophysical transients detectable by CZTI is low enough that nearby orbits are unlikely to have a large number of transients.

We first decide the width of the window used for transient search, say $t_w = 100$~s, and an acceptable \edited{false positives probability (FPP, $\mathcal{F}$)}. We typically set $\mathcal{F}=0.1$ for a single quadrant. Since we place limits using data from all four independent quadrants, the combined FPP is $10^{-4}$. We now need to find a ``cut-off rate'' $R_c$ such that the probability of this threshold being crossed by chance in $t_w$ is $\mathcal{F}$. To calculate $R_c$, we select five orbits before and after the transient (excluding the orbit containing the transient) as ``witness'' orbits. We create light curves for these orbits using the same time bin as used in the original analysis, then de-trend them, and create histograms of the de-trended counts. $R_c$ is defined as the point such that a fraction $\mathcal{F}$ of the data points have counts $>R_c$. A typical orbit has 4000~s-5000~s of usable data, so that analysis of ten orbits with parameters $\mathcal{F}=0.1$ and $t_w = 100$~s ensure that 40-50 data points are above $R_c$. This makes the method robust to the presence of another transient in the witness orbits.

There are some caveats to be noted here. Occasionally, a quadrant can be extremely noisy in some orbit. If the candidate transient is in such an orbit, that quadrant is excluded from further analysis and there is a corresponding decrease in the FPP \citep[for instance][]{2017GCN.20796....1M,2019GCN.24972....1M}. Our \edited{FPP} estimates are derived from the probability of getting counts $>R_c$ in each of the four quadrants anywhere in the $t_w$ window. In practice, we consider something a detection only if such spikes in counts are coincident across multiple quadrants, hence the actual \edited{FPP} is even lower.

\subsection{Flux calculations}
Incident photons from off-axis transients are heavily re-processed (scattering, absorption, fluorescence, etc) by various satellite elements before they are incident on the detector. Hence, the mapping of incident spectra to measured spectra must be done by simulating these effects in software. We accomplish this by using a GEANT4-based mass model of the entire satellite (Mate et al., this volume). Since the effect of the satellite varies with direction, the simulations require knowledge of the source position in satellite coordinates. For transients where the position is known, Chattopadhyay et al (this volume) discuss a method of estimating the source spectrum and flux from CZTI data.

While methods for calculating the source spectrum are still under development, we have found that source flux calculations based on the mass model are quite reliable if the source spectrum is known from other instruments. We leverage this by assuming a power-law or band model spectrum for sources, and calculating the flux corresponding to the number of counts in a quadrant. The total flux from the source is the sum of fluxes in all four quadrants.
%The total upper limit on the source flux is the sum of flux limits for all four quadrants \citep[see for instance][]{2020ApJ...888...40A}.

For certain transients, most notably gravitational wave events, the source location is not known precisely. Instead, discovery teams provide a sky-map with the source position probability distribution. For such sources, we evaluate the flux limit at each point on the sky map that is not occulted by the Earth at the instant of the transient. The overall flux limit is evaluated as a probability-weighted mean of these values \citep[for instance see][]{2020GCN.27315....1S}.

\section{Blind searches for transients}\label{sec:search}
%\rough{just mention ML code, cite paper, then talk about stat-based searches}
The triggered searches are complemented by a broad ``blind'' search over all of CZTI data to identify astrophysical transients. We have two pipelines for such searches --- a pipeline based on machine learning (ML) \citep{Abraham2019} and the CIFT\footnote{CIFT is pronounced as sift.}. In this section, we discuss CIFT in detail.

The broad outline for the CIFT searches is as follows: First, data are reduced and de-trended as discussed in \S\ref{sec:dataprep}. Various algorithms are used to identify outliers in light curves. These outliers are used to create `peak maps' to identify candidate transients in data. Flagged candidates are displayed on an interface for human vetting. They undergo similar quality checks and inspection as discussed in \S\ref{sec:triggered}, and final selected transients are saved in a database.
%Next, we discuss the three search algorithms currently used: Top-N, N-sigma, and cut-off rates.

\subsection{Preparing the data}\label{sec:bandsmaps}
CZTI Level 2 bunch cleaned files are organised into `Obs-ID's which have all the data taken during observations of any particular object requested by an observer. We undertake most of our searches Obs-ID wise, thus typically processing a few to a dozen orbits at a time.
We see that noise events are more frequent in lower energies, while data are cleanest at higher energies. To leverage this factor, we divide CZTI data into three energy bands: 20--50~keV, for 50--100~keV, and 100--200~keV. For all three bands, we process the data following steps from \S\ref{sec:dataprep}, and create de-trended light curves with 0.1~s, 1~s, and 10~s bins. We also use a 0.01~s binning when searching for counterparts to fast radio bursts. We use the entire energy range for the Veto detector, and create light curves at 1~s and 10~s binning.

Thus, we generally create 36 light curves for CZTI data (3 time bins $\times$ 3 energy bands $\times$ 4 quadrants) and 8 light curves for Veto data (2 time bins $\times$ 4 quadrants) per Obs-ID. We run a search algorithm on each light curve to identify outliers and create `peak maps': boolean masks with value 1 for time bins containing the outliers, and 0 elsewhere. The twelve CZTI peak maps are added together, and any bin with a mask value of four or higher is flagged as a candidate transient. Similarly, the four veto masks are combined and bins with mask value $\geq 3$ are flagged as candidate transients. Next, we discuss the three outlier search algorithms currently implemented in CIFT.
%Peak maps are added together and any bin with a mask value of 4 (3) is a candidate transient (effectively meaning coincident detection in at least 2 quadrants for CZTI and 3 for Veto).

\subsection{Top-N}\label{sec:topn}
The Top-N (TN) algorithm is based on a simple heuristic: a transient is expected to have among the highest count rates seen in a given light curve. We identify the brightest $N$ bins in a light curve and flag them as outliers for the peak map.
%In its simplest form, we should simply check if the brightest bins in the lightcurves for the four quadrants are coincident with each other. This becomes overly restrictive, as the transient may have peaked close to a bin boundary and as such the brightest bins are consecutive rather than coincident. It is also too sensitive to Poisson variations in the transient lightcurve. To help overcome these hurdles, we identify the brightest $N$ bins in the 12 lightcurves corresponding to three energy bands and four quadrants. The boolean masks are created, set to 1 for the $N$ bins and 0 for the rest of the bins. We then add the masks for all 12 lightcurves. Any bin with a mask value of 4 is a candidate transient (effectively meaning detection in at least two quadrants).

While testing this algorithm, we obtained better results if the searches were carried out one orbit at a time (as opposed to Obs-ID wise searches for other algorithms). By varying values of $N$, we obtained the best results for $N = 3$.

%After testing algorithm performance, we decided to run Top-N algorithm on orbit-wise lightcurves since it is optimal for small number statistics while N-sigma and cut-off rate algorithms are run on combined lightcurves of all orbits in an Obs-ID since they perform better with large data sets. }

% \todo{To be clarified: is this search done in 12 bands, and a match of 4 is required?}
% \rough{Aditi: A match of 4 or above is considered a detection}

% \todo{To be clarified: at one point, we used to ``grow'' the mask such that top-n (or others) in adjacent bins were also considered coincident. Is that still the case?.}
% \rough{Aditi: No, we no longer ``grow'' the mask (confirm with Yashvi)}
% \rough{We never used to grow the mask.}
%By varying values of $N$, we found that $N = 3$ is most effective for CZTI data.

\subsection{N-sigma}
The N-sigma (NS) algorithm is a straightforward statistics-based method to select outliers in a time series. We identify outliers by using iterative sigma clipping as implemented in the \sw{Astropy} \sw{sigma\_clipped\_stats} module. Starting with a de-trended light curve, we calculate the median and standard deviation ($\sigma$) values, and reject outliers that deviate more than 3$\sigma$ from the median. The process is repeated with the new light curve until convergence is attained, subject to a maximum cap of five iterations. The mean value $\mu$ and the standard deviation $\sigma$ of the final iteration become the key parameters of algorithm. Using these values, outliers are defined as data points with counts $ > \mu + N\sigma$, where our default value is $N = 5$. The typical thresholds for flagging these outliers for various time bins, energy bands, and both detector types are given in Table~\ref{tab:combinedcut-off}. These values were calculated from data of entire five years of the search. We reiterate that the namesake $N$ of this method is used only in identifying outliers for the peak map, while the iterative sigma estimation is always done at a three-sigma level.

%We apply sigma clipping (using Astropy's \footnote{https://www.astropy.org/about.html} \texttt{sigma\_clipped\_stats} module) to calculate the mean, median and standard deviation of background noise. Sigma clipping works as follows; the median(m) and standard deviation($\sigma$) of the data are calculated and all points outside the range $ m \pm  \mathrm{N} * \sigma $ are rejected where N is specified by the user. The choice of median over mean is governed by the fact that medians are more robust to outliers. This process is repeated until one of the exit criterion are met i.e. either user specified fixed number of iterations are over, or the fractional change in standard deviation converges. This method doesn't perform well on small data sets as reliability of median decreases. It is most useful when the signal doesn't cover more than half of the full data set.

%We then define the cutoff rate as
%$$ \mathrm{Cutoff \hspace{0.2cm} rate} = \mu + \mathrm{N} * \sigma $$
%where $\mu$ is the mean and $\sigma$ is the final standard deviation of underlying noise. Boolean masks are created where bins with rates above this cutoff rate are set as 1 and rest as 0 for all lightcurves.

\begin{table}[!h]
    \centering
    \begin{tabular}{c c r c l}
         \toprule
	\toprule
	{Method}& {Binning} & {} & {Band-wise cutoff} & {}\\
	{} & {(s)} & 0 & 1 & 2\\
	\midrule
    \textbf{CZTI} & & & &\\
	\midrule
	&0.1&1954&488&428\\
	{Cutoff rate}&{1.0}&263&107&102\\
	&{10.0}&28&22&22\\
	\midrule
	&{0.1}&396&410&407\\
	{NSigma}&{1.0}&137&133&131\\
	&10.0&41&38&38\\
	
	\toprule
	\toprule
	{Method}& {Binning} & {} & {Combined} & {}\\
	{} & {(s)} &  & cutoffs  & \\
	\midrule
	\textbf{VETO} & & & & \\
	\midrule
	\multirow{2}{*}{Cutoff rate}&1.0& & 319 & \\
	&10.0& &690 & \\
	\midrule
	\multirow{2}{*}{NSigma}&1.0&&394&\\
	&10.0&&1150&\\

    \end{tabular}
    \caption{Combined cut-offs for Cutoff rate and NSigma methods for each binning and band. These representative rates were calculated by using all the the five years of data used in this study. Note that rates are in units of counts/sec, not counts per bin.
\label{tab:combinedcut-off}}
\end{table}

\subsection{Cutoff rates based on False Positive Probability}
The cutoff rate based search (CR) algorithm aims at attaining a given \edited{False Positive Probability (FPP)} for candidate transients. Cutoff rates are determined following the procedure discussed in \S\ref{sec:nondet}, with one important distinction. In \S\ref{sec:nondet}, we assumed the presence of transient-free data of an order of magnitude larger duration than the timespan of interest. Since CIFT searches are meant to be conducted over all available data, this requirement clearly cannot be met. Instead, we set our FPP threshold based on the expected rates of transients, in particular, GRBs.

The rate of detectable GRBs is a function of instrument sensitivity, energy range, and field of view. As a baseline, we note that on average {\em Fermi} GBM detects a GRB every 1.5 days \citep{2020ApJ...893...46V}, while the BAT on the {\em Neil Gehrels Swift Observatory} averages one GRB every four days \citep{2016ApJ...829....7L}. Based on these we stipulate a rough upper bound of the rate of GRBs detectable by CZTI as 0.5 GRBs per day\footnote{We note that the subsequent arguments become stronger if the actual detected rate is lower as was expected. After completing the search, indeed we found a much lower GRB rate.}. We then stipulate that only 1\%\ of our GRBs may be \edited{false positives (FPP = 0.01), corresponding to one false positive every 200~days.}

%% move table here %%

To arrive at an approximate solution for the \edited{FPP} criterion, we consider the case of searching for a GRB with 1~s duration in light curves with 1~s binning. In this scenario, our \edited{false positive} requirement of 1 per 200~days maps to one false positive in $1.728 \times 10^7$ bins. Since most basic acceptance criterion is coincident detection in two or more independent quadrants, each quadrant can have one \edited{false positive} in $\sqrt{17280000} $ time bins, or 4156~s. This is a significant fraction of an orbit, and hence the robust estimation of $R_c$ requires data from several orbits. Decreasing the time bin size increases the number of samples in the light curve, and owing to the random underlying process, makes outliers more likely. To correct for this, we change our cutoff rate requirements based on the bin size $t_\mathrm{bin}$: $R_c$ is selected such that a fraction $0.01 \times (t_\mathrm{bin} / 4156~\mathrm{s})$ of bins have a count rate $> R_c$.

We note that this is a highly simplified argument, which ignores the 12 light curves we make for every time bin and the $>4$ peak map condition. It also ignores the small effect of presence of transients in our ``witness'' data sets. However, it serves as a good approximate argument for selecting our $R_c$ thresholds from data.

The typical thresholds for flagging these outliers for various time bins, energy bands, and both detector types are given in Table~\ref{tab:combinedcut-off}. As in the NS method, these representative rates shown in the table were calculated all five years included in this work. Some specialised searches use the entire 20--200~keV range as a single band. For such searches with 1~s binning, the cutoff rates for the 4 quadrants are 79, 68, 68, and 69 counts/sec respectively. For searches with 10~s binning, the rates drop to 10, 10, 10, and 12 counts/sec respectively, corresponding to a total of 420 counts per 10~s bin.

%At this stage, we have to choose how to map this FAR requirement to other time bin sizes: we can either state the FAR as number of false alarms per time, or number of false alarms per light curve bin.
%Furthermore, we opt to define the FAR as a function of number of bins searched given by the ratio of the timespan to the bin size.

%In this method, confidence expected from a given false alarm rate (FAR) is calculated and equated to confidence from histogram of count rates to get cutoff rate. This algorithm is currently tailored for GRBs. On average, we expect to see one GRB per one or two days, i.e. 1 GRB in 2x86400 seconds. For a FAR of 1 per 100 GRBs, it gives 1 FAR per 17280000 seconds. This value corresponds to all detections and since we want to check for coincidence in at least two quadrants, we get a FAR of 1 per $\sqrt{17280000}$ seconds per quadrant. This `Timespan' turns out to be approximately 4156 seconds. Hence,
%$$ \mathrm{Confidence} = 1 - \frac{\mathrm{bin size}*\mathrm{FAR}}{\mathrm{Timespan}} $$
%where binsize refers to temporal binning.
%
%From histogram of each quadrant,
%$$ \mathrm{Confidence} = \frac{\mathrm{Cumulative \hspace{0.2cm} Sum}}{\mathrm{Total \hspace{0.2cm} Sum}} $$
%Equating the two above, we get cutoff rates for all 12 lightcurves and use them to create boolean masks. Since an Astrosat orbit is roughly 6000~s, with a Timespan of 4156~s as calculated above, this algorithm would miss weak GRBs if the search is done orbit-wise hence this algorithm is run on combined lightcurves of all orbits in an Obs-ID.

\subsection{The CIFT interface}
Once the peak maps have been created by any of the three algorithms discussed above, we apply our candidate selection criteria of requiring $\geq 4$ matches out of 12 light curves for CZTI, and at least three matches out of four veto light curves (\S\ref{sec:bandsmaps}). Candidate transients that meet this requirement are flagged as an ``event'', and entered into an SQL database. Certain basic properties like like number of quadrants and energy bands an event was detected in, their significances, rates above background, time since last SAA, time from next SAA, etc are also calculated and stored in the database. Events having the same trigger time (for instance if they were detected by two different algorithms) are grouped, and their corresponding
event-IDs are stored under a unique trigger-ID in a separate table. Furthermore, the trigger-events which are within 100 seconds of each other are grouped into a ``superevent'' and assigned a super-ID. These superevents are the final transient candidates, ready for human inspection.

%The time and count rate in these outlier bins (or transient candidates) is noted and an `event' is created for each candidate. Basic statistics, like number of quadrants and energy bands an event was detected in, their significances, rates above background, time since last SAA, time from next SAA, etc are calculated for each event. In addition for CZTI, each energy band is assigned a weight : 1 for 20--50~keV, 2 for 50--100~keV and 3 for 100--200~keV. This weight is multiplied by the number of quadrants the peak was detected in that band and summed. This parameter is referred as `rank' and is an estimate of how strong the event was as seen by CZTI. It is used to judge the reliability of an event.

A separate program for plotting is run in parallel which takes input a list of Obs-IDs and fetches all the superevents in those Obs-IDs from the SQL database. For each superevent, it plots detailed time energy histograms, light curves and calculates \tnt\ for each temporal binning.

\begin{figure*}[htp]
\centering
\begin{subfigure}{0.6\linewidth}
\centering
	\includegraphics[width=\columnwidth]{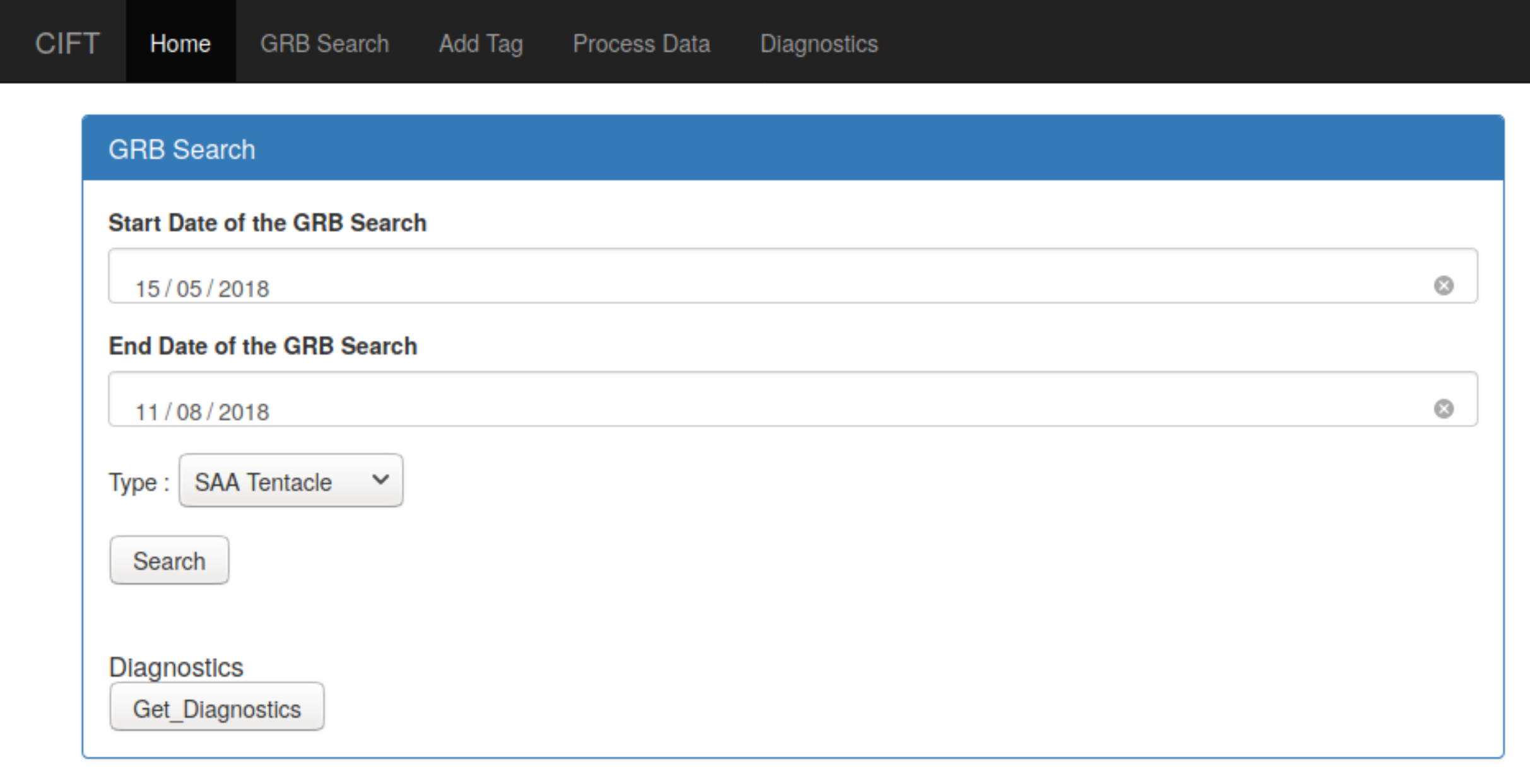}
	\caption{Main page}\label{fig:cift_main}
\end{subfigure}

\begin{subfigure}{0.6\linewidth}
\centering
	\includegraphics[width=\columnwidth]{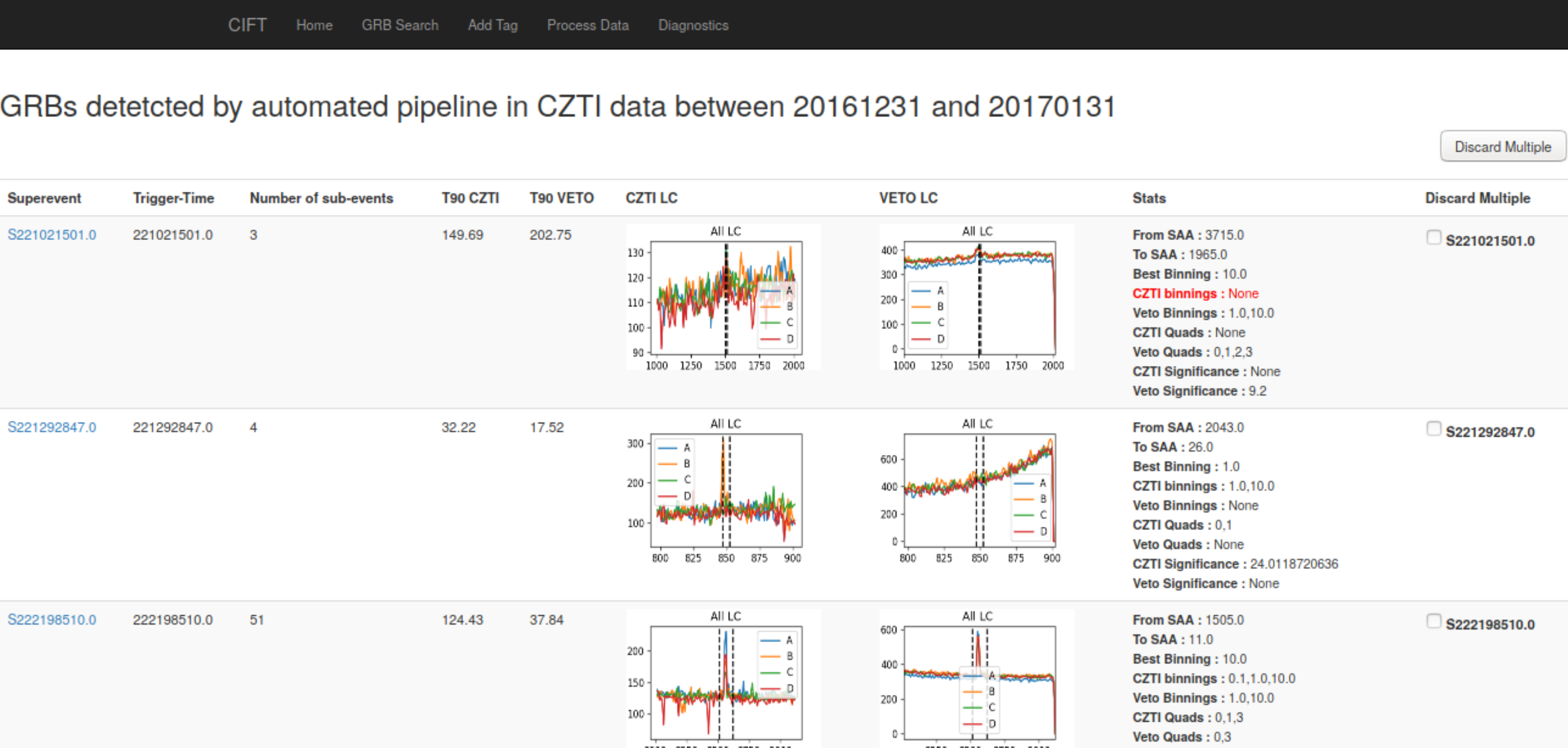}
	\caption{Scanning page}\label{fig:cift_scan}
\end{subfigure}

% \begin{subfigure}{0.49\linewidth}
% \centering
% %	\includegraphics[width=\columnwidth]{SpectrogramNorm}
% 	\caption{Superevent view} \label{fig:cift_view}
% \end{subfigure}
\begin{subfigure}{0.6\linewidth}
\centering
	\includegraphics[width=\columnwidth]{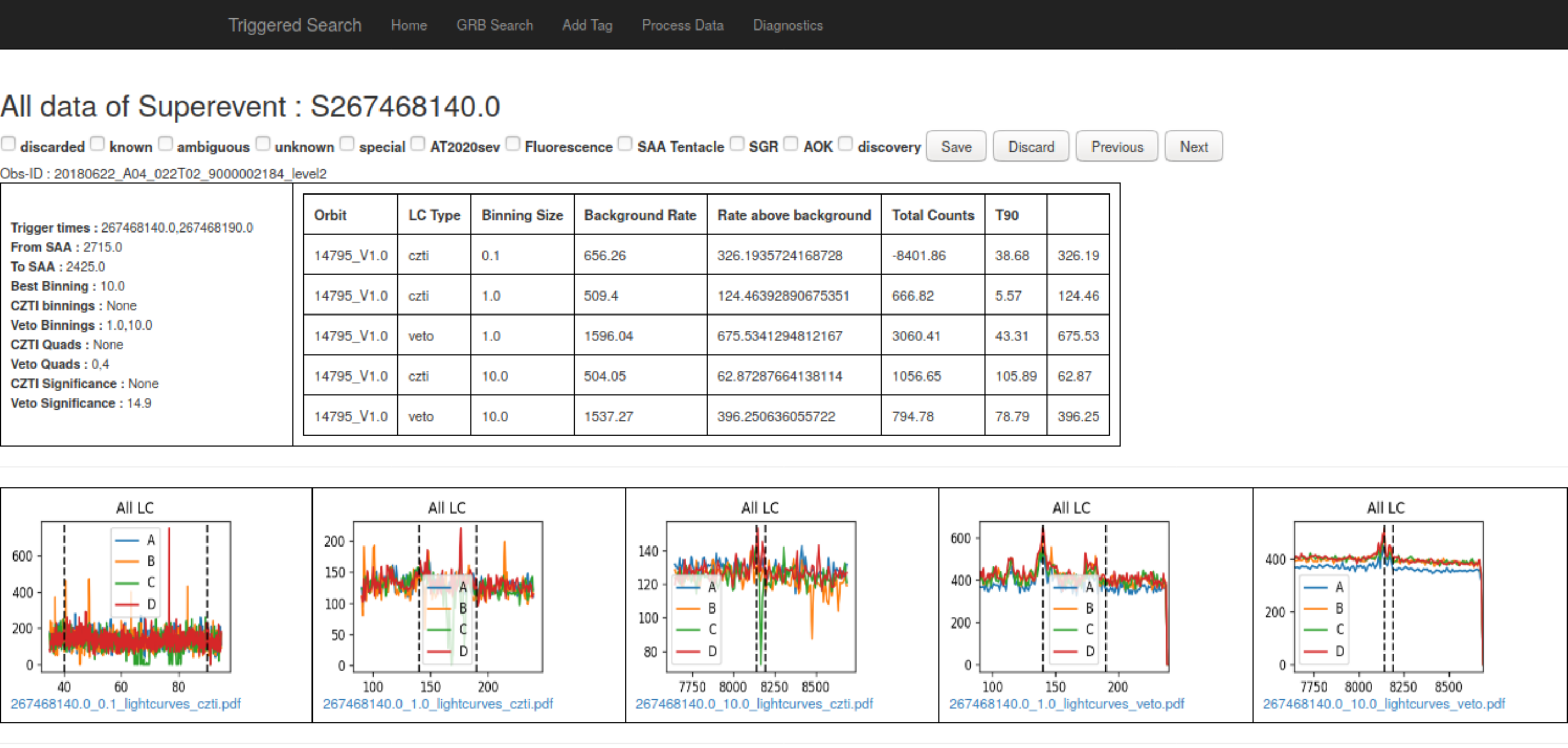}
	\caption{Inspection page}\label{fig:cift_inspect}
\end{subfigure} \hfill

\caption{Screenshots of CIFT, showing functionality of various pages. Panel \subref{fig:cift_main}: The screenshot of the home page of CIFT, where the human scanner can input dates and the corresponding candidate tag which the scanner wants to see, refer Section 4.5. This main page also allows the user to navigate to other functionalities of the interface where one can add new tags, process the available unprocessed directories and access the diagnostics page, with few clicks.  Panel \subref{fig:cift_scan}: The SQL database displays all the candidates of the specified tag from all Obs-IDs contained within the date range specified on the CIFT main page. Each candidate has a dedicated row where the Superevent-ID, trigger time, $T_{90}$ in both CZTI \& Veto, number of sub-events are displayed along with relevant statistics like background rate, peak rate, number of quadrants where the candidate was detected for quick reference of the scanner. Each row also has a appropriate lightcurve thumbnail for both CZTI \& Veto for visual inspection, allowing the scanner to discard the very obvious bogus candidates from this page itself (with the help of discard multiple option).  This complete list of candidates is sorted in the ascending order of the Superevent-ID. Panel \subref{fig:cift_inspect}: Each candidate is linked to their inspection page which displays the break-down of all the computed characteristics shown on the Scanning page. The inspection page also contains links to five different lightcurves for different binnings of CZTI and Veto detectors. Based on the inspection of all these parameters and lightcurves, the scanner can classify the candidate and tag the candidate with the appropriate tag. }
\label{fig:cift}
\end{figure*}

The CZTI Interface For Transients (CIFT) is a Flask\footnote{https://pypi.org/project/Flask/}-based interface with SQL database as back-end, available to view the candidate transients (Figure~\ref{fig:cift}).
%The interface can be accessed remotely using \sw{ngrok} \footnote{https://ngrok.com/}.
The interface allows a human scanner to search all superevents by Obs-ID which are displayed in a table on a `scanning' page (Figure~\ref{fig:cift_scan}). The scanning page has columns for Superevent-ID, trigger time, \tnt\ in CZTI and Veto, number of sub-events, a column displaying relevant statistics like background rate, peak rate, number of quadrants the candidate was detected in, etc. and a check-box option to discard multiple superevents at once if bogus. Each superevent-ID is linked to an inspect page (Figure~\ref{fig:cift_inspect}) which lists all characteristics of the superevent, and of each sub-events contained within it, along with several lightcurves of different binning sizes for CZTI \& Veto. After inspection, a human scanner can tag the event with custom tags, including ``known'', ``unknown'', ``ambiguous'', ``SAA Tentacle'', etc. Superevents can be searched and filtered by tags from the main page (Figure~\ref{fig:cift_main}). The CIFT interface also has other features like undertaking triggered searches and a front-end for initiating data processing.\\
% \rough{Several parameters/characteristics used on the inspection page aren't listed here, should they be?}\rough{ Aditi: Band and Quadrant-wise ``Rank'' and ``significance'' for each detection event. T90, background rate, rate above background and total counts for each of the binnings in CZTI and VETO}
% \rough{Varun: only if they are important. give me a list}
%are linked to pages to view and inspect superevents which redirect to an `inspect' page containing detailed information about sub-events of a superevent. After inspecting the superevent in detail, the user can assign a tag from a list of options like discarded, known, unknown, ambiguous, SAA Tentacle, to name a few. The user can also create a new tag using the `Add Tag' option in the navigation bar. The main interface page also allows for filtering superevents by assigned tag for easy viewing of real transients. The interface can also be used to start data processing using the `Process Data' tab apart from command line. This page lists the directories available to process and the user can select multiple directories at once using check-boxes.
%\todo{Add screenshots of the Interface Pages}\rough{can go in appendix}
%\rough{Aditi: Mention about the interface for Triggered Searches as well}

\section{Results}\label{sec:results}
%\rough{total = sum of bottom row of table 1}
% We detected a total of 227 transients in CZTI data by using CIFT. Of these, 36 are GRBs or triggers previously reported by other missions (\S\ref{sec:pocmiss}), while 27 are new transients (\S\ref{sec:new}). In the same five-year span, triggered searches and the ML pipeline have detected 330 GRBs, of which our searches recovered 270. \todo{add the missing slew orbit comment here and in the missed GRB section.}The reasons for the missed GRBs are discussed in \S\ref{sec:missedGRB}.
We used our framework to search for GRBs in data from 06 October 2015 when CZTI was first powered on, till 10 October 2020 --- spanning just over five years of data.
``Slew'' Obs-IDs are relatively short data sets acquired when \asat\ is slewing from one source to another. These have been excluded from our search. We detected a total of 348 transients in CZTI data by using CIFT. Of these, 41 are GRBs or triggers previously reported by other missions but missed by POC triggered searches or ML pipeline (\S\ref{sec:pocmiss}), while 47 are new discoveries (\S\ref{sec:new}). In the same five-year span, triggered searches and the ML pipeline have detected 325 GRBs, of which our searches recovered 260. Two of these missed GRBs were in slew orbits. The reasons for missing $\sim$20\% GRBs are discussed in \S\ref{sec:missedGRB}.
%We had not previously processed the data taken during slewing (slew orbits) but during analysis of missed GRBs, we discovered \todo{2 of them} in slew orbits, and hence the processing of all remaining slew orbits is now being undertaken. The reasons for the missed GRBs are discussed in \S\ref{sec:missedGRB}.

\subsection{Performance}
The processing code takes less than an hour to search for transient candidates in one month of data (approximately 130~GB). Creating diagnostic plots is a slower process which is spawned in parallel, and takes 3--4 hours to complete. Users remotely connect to the http-based interface for scanning the processed data. Visual examination of candidates from a month of data takes a few hours for an experienced user.

%\rough{commented out the 3-year text}
%\rough{All the numbers are for the 3 years of 2017, 2018, 2019.}
%From the archival analysis of all data since AstroSat's launch in 2015 till September 2020, CIFT has discovered a total of 27 new GRBs. Out of 190 GRBs detected by POC pipeline in CZTI data\footnote{http://astrosat.iucaa.in/czti/?q=grb}, CIFT detected 164, missed 26 and detected 63 new GRBs which were missed by POC pipeline, the detailed analysis of which is explained in \S\ref{sec:missedGRB}. Table \ref{tab:Summary Table} summarizes number of total CIFT candidates, detected by both POC and CIFT, detected by CIFT and not POC, and new discoveries by CIFT for the duration October 2015 to 10 October 2020.

% Figure Xa shows percentage of known GRBs, new GRBs and bogus discarded superevents. Figure Xb shows distribution of all real GRB detections (both known and new) across CZTI, Veto data and three algorithms. Figure Xc shows the same distribution for discarded superevents.\\

\begin{figure*}
 \centering
 \begin{subfigure}{0.5\textwidth}
 \includegraphics[width=\textwidth]{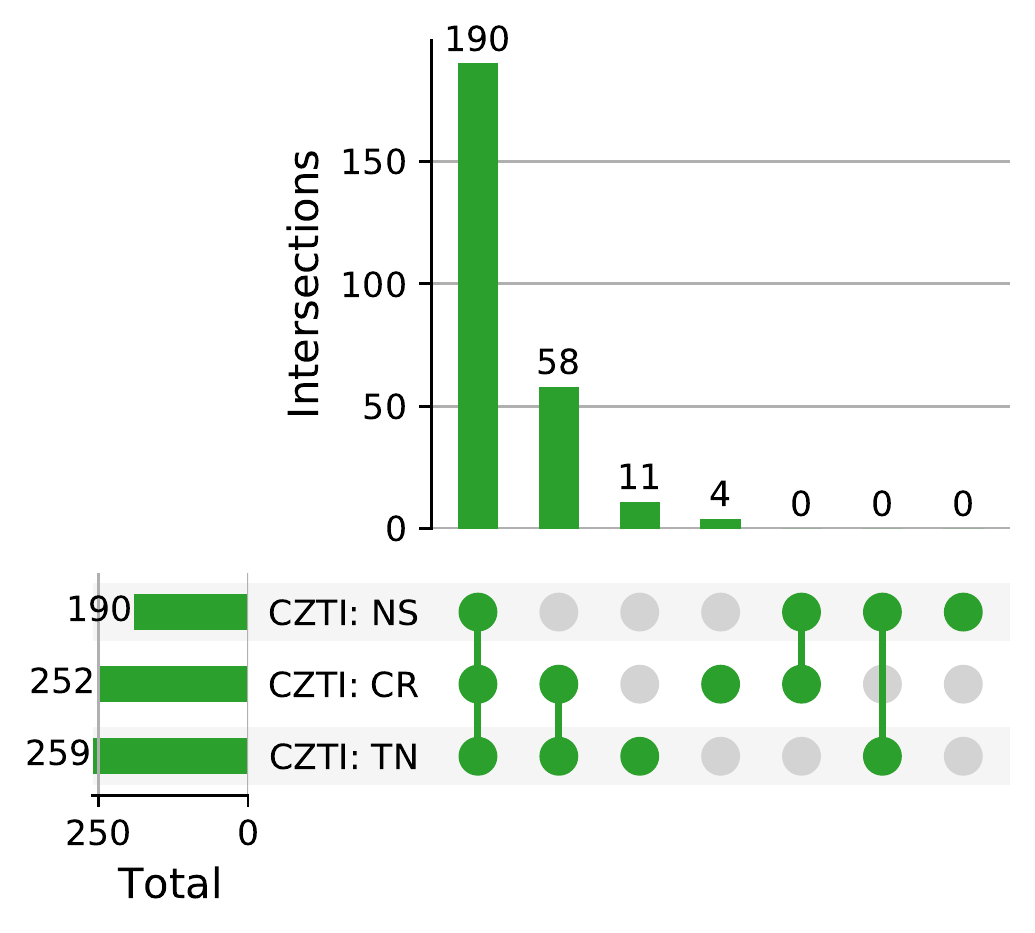}
 \caption{Statistics of transients detected in CZTI.}
 \end{subfigure}
 ~
\begin{subfigure}{0.5\textwidth}
 \includegraphics[width=\textwidth]{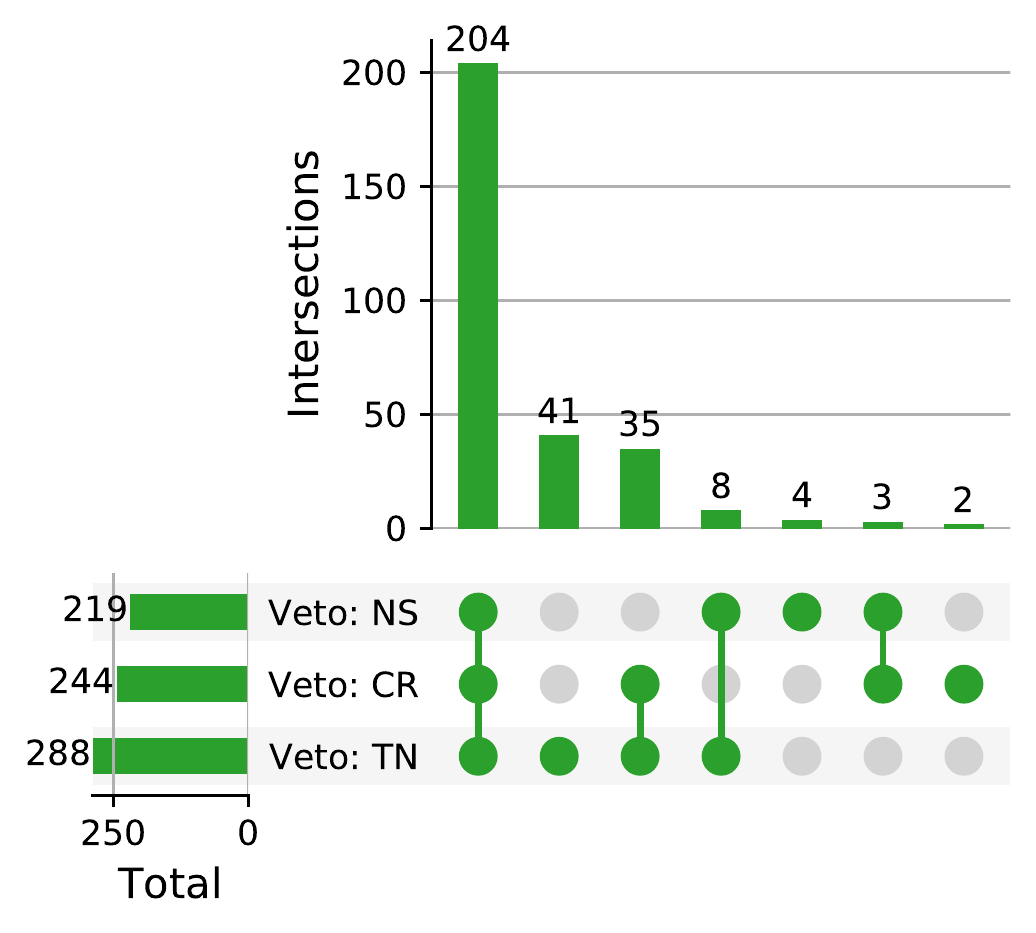}
 \caption{Statistics of transients detected in Veto detectors.}
 \end{subfigure}
\caption{Statistics of transients detected in CZTI (Panel a) and Veto (Panel b) detectors. The three bottom rows show the number of transients detected by each method: for instance, the N-sigma method detected 115 transients in CZTI. The bar charts at the top show overlaps between various combinations of methods. We see that of the 263 transients detected by CZTI detectors, 190 were detected by all three methods, while another 58 were detected by both the ``cutoff rates'' and ``Top-N'' methods. Among the 297 transients found in the Veto detector, 204 transients were detected by all three methods, 41 were detected only by the ``Top-N'' method, and 35 were detected by both the ``cutoff rates'' and ``Top-N'' methods.}
\label{fig:upset_CZTI}
\end{figure*}

Figure~\ref{fig:upset_CZTI} shows the break-up of transient detections by the various algorithms. We see that most transients are detected by all three algorithms, followed by detections in both CS and TN. The TN method is solely responsible for the detection of 15\% of Veto transients.

Table~\ref{tab:summary} summarises the performance of all algorithms. We see that there are a large number of false positives, particularly from the Veto detectors. This underscores the need for human vetting of the candidate superevents. 

On an average, CIFT flags about 339 candidates per month, adding up to 19628 candidates in 58 months of data. For the months of April and May 2020, we lowered the thresholds to search for even faint bursts associated with the outburst of the galactic magnetar / FRB candidate SGR~1935+2154 \citep{2020ApJ...898L..29M}. We selected the top 5 peaks in the TN method, and required a coincidence of just 2 bands out of 12 in CR and NS methods. These reduced thresholds increased the number of candidates by a factor of 4.3, giving 2936 candidates in just 2 months.
% Out of 22564 candidates throughout the 48.3 months that were processed and scanned, April and May of 2020 were processed with lowered thresholds to study SGR 1935+2154 bursts more closely. The average number of candidates in these months is 1468 per month. The average number of candidates excluding the above mentioned months is 425.

The most common type of false positives comprised of coincident detections in two Veto quadrants in just a single second, with no discernible signal in adjacent bins. These are most likely particle events, and are rejected. A closely associated class of veto false positives are events that have a very sharp rise and an exponential decay: again a profile common for particle events. On the other hand, Veto light curves of GRBs that are also detected in CZT detectors show a wider variety. Hence we decided to keep the coincidence threshold for veto as 3 out of 4 quadrants at the expense of missing possible real transients, and this was the number discussed at the start of \S\ref{sec:search}.

Other large number of bogus detections include false peaks near SAA due to bad de-trending or inadequate SAA masking which can be ruled out during human vetting. In CZTI data, many false events are caused by a single pixel, generating noise events at all energies. Visual examination of the distribution of counts in the detector helps to quickly dismiss these as false positives. If the light curves are well-behaved with no real transients or noise spikes, then the TN algorithm often generates false positives by identifying ``outliers'' that are completely consistent with background.

As human scanners gain more experience with the pathologies of false positives, we are working to improve automatic rejection of such candidates.
% \todo{any other types of false positives and how they are rejected} \rough{ Aditi: 2 CZTI quadrants single lower energy bin, Single quadrant flickering pixel visible in all energy ranges}

%For real time processing, CIFT would take only an hour to process and plot single Obs-ID data, which could be scanned by a human in less than an hour to publish detected real transients everyday.

\begin{figure*}[!hp]
 \centering
 \begin{subfigure}{\textwidth}
    \centering
    \includegraphics[width=0.85\textwidth, height=0.45\textheight]{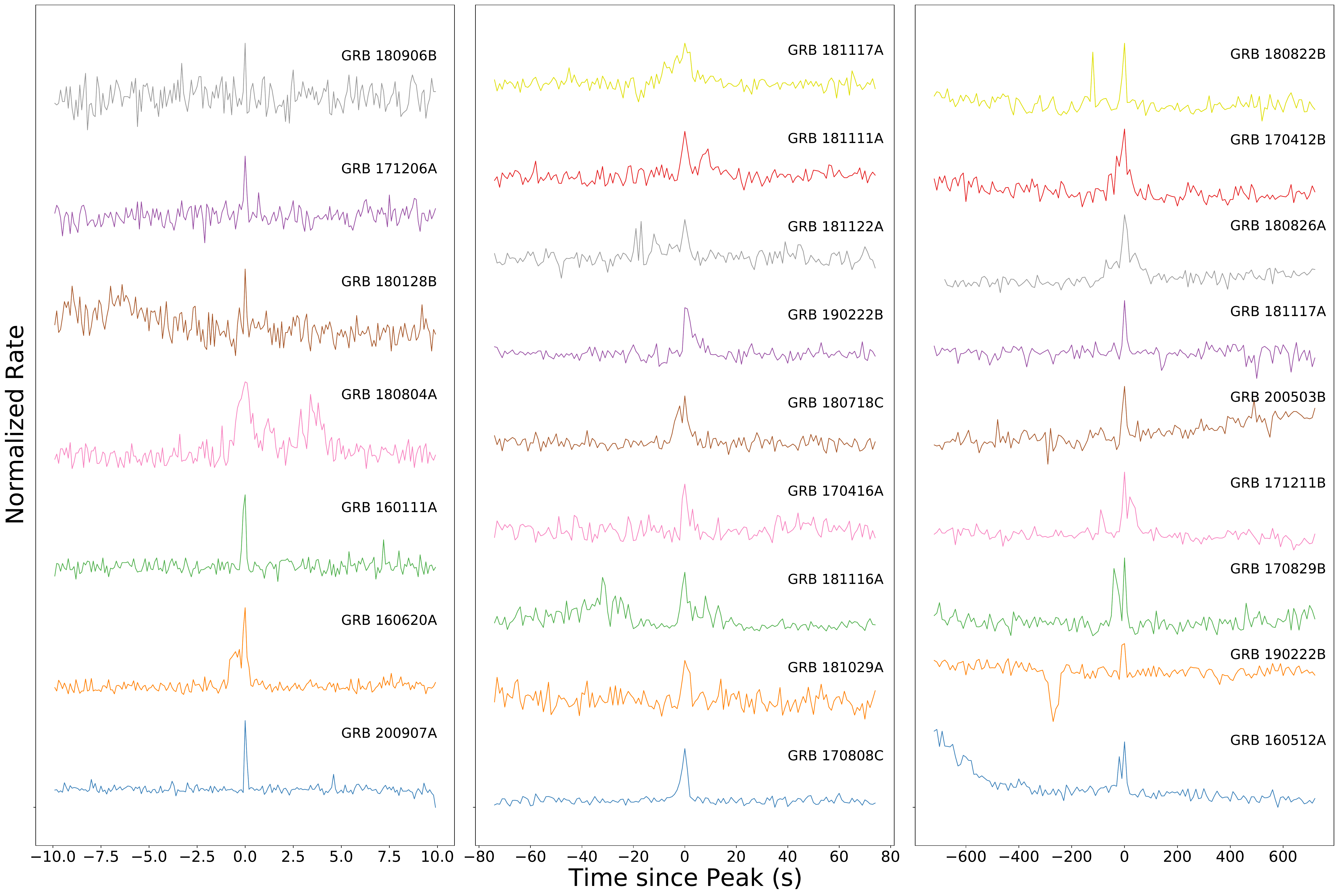}
    \caption{CZTI lightcurves}
    \label{fig:unknown_lc_czti}
 \end{subfigure}
  ~
  \begin{subfigure}{\textwidth}
    \centering
    \includegraphics[width=0.85\textwidth, height=0.45\textheight]{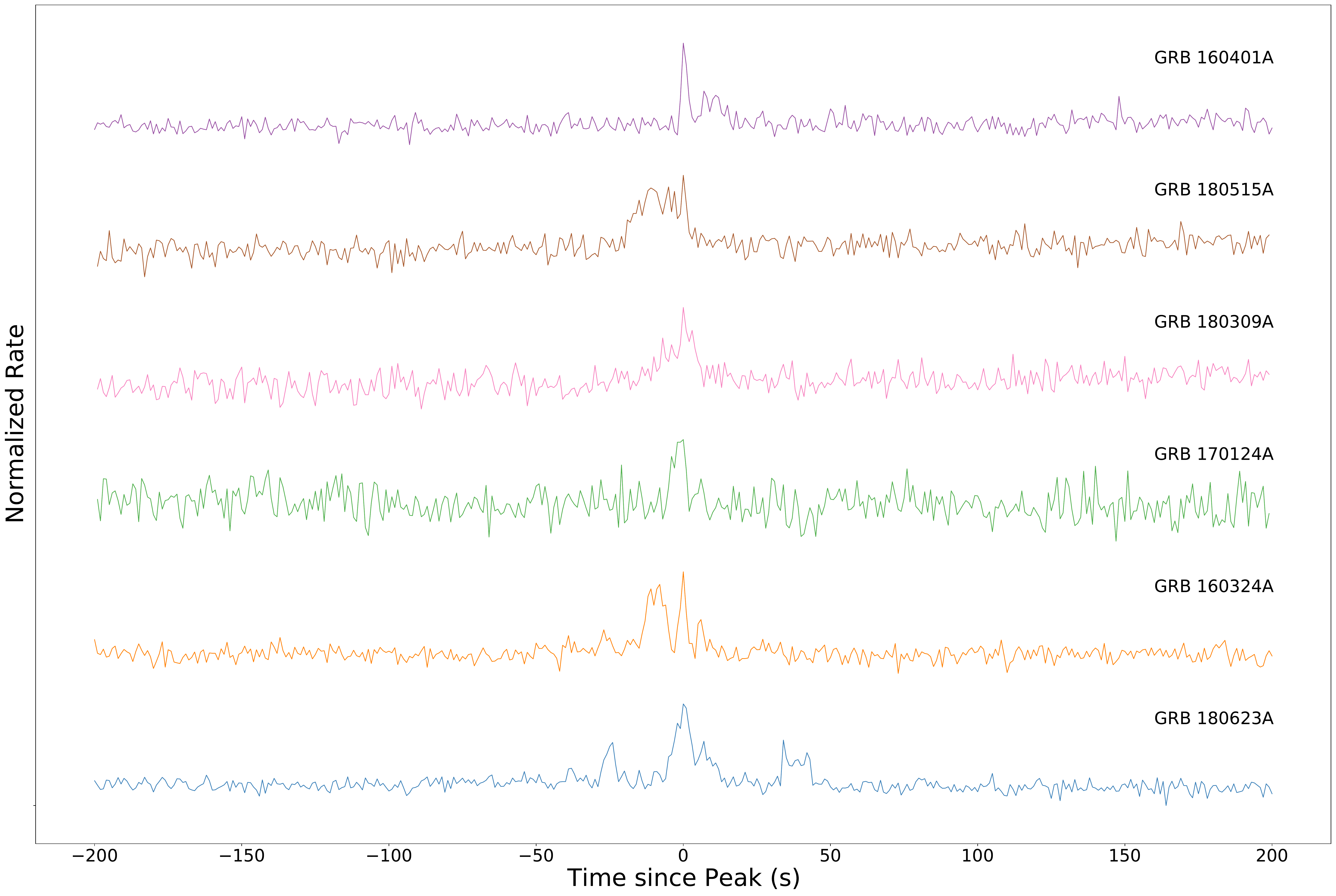}
    \caption{Veto lightcurves}
    \label{fig:unknown_lc_veto}
 \end{subfigure}
    \caption{The normalised lightcurves of GRBs detected by CIFT, that were reported by other instruments but had not been identified in CZTI or Veto data (\S\ref{sec:pocmiss}). Each GRB light curve is normalised and labeled with the GRB name. Panel \subref{fig:unknown_lc_czti} shows normalised lightcurves for the GRBs detected in CZTI. The three sub-panels are with 0.1~s, 1~s and 10~s binning respectively, and each sub-panel is ordered by peak count rate above background, increasing from top to bottom. Panel \subref{fig:unknown_lc_veto} shows the normalised lightcurves of GRBs that were detected only in Veto. These are plotted with a 1~s binning, and are also ordered by peak count rate above background, increasing from top to bottom.}
    \label{fig:knownGRBlc}
\end{figure*}

\begin{figure*}[!hp]
 \centering
 \begin{subfigure}{\textwidth}
    \centering
    \includegraphics[width=0.85\textwidth, height=0.45\textheight]{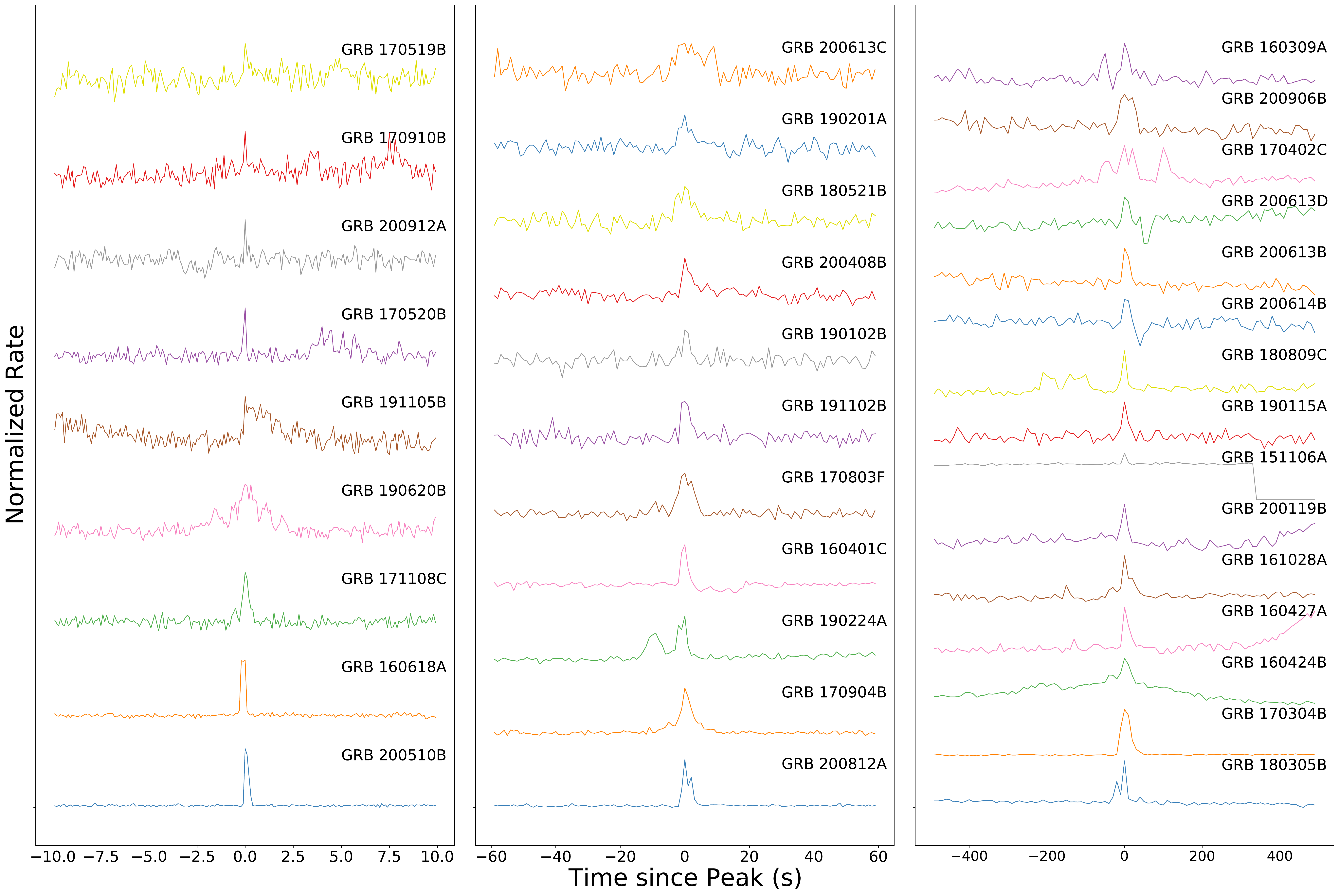}
    \caption{CZTI lightcurves}\label{fig:discovery_lc_czti}
 \end{subfigure}
  ~
  \begin{subfigure}{\textwidth}
    \centering
    \includegraphics[width=0.85\textwidth, height=0.45\textheight]{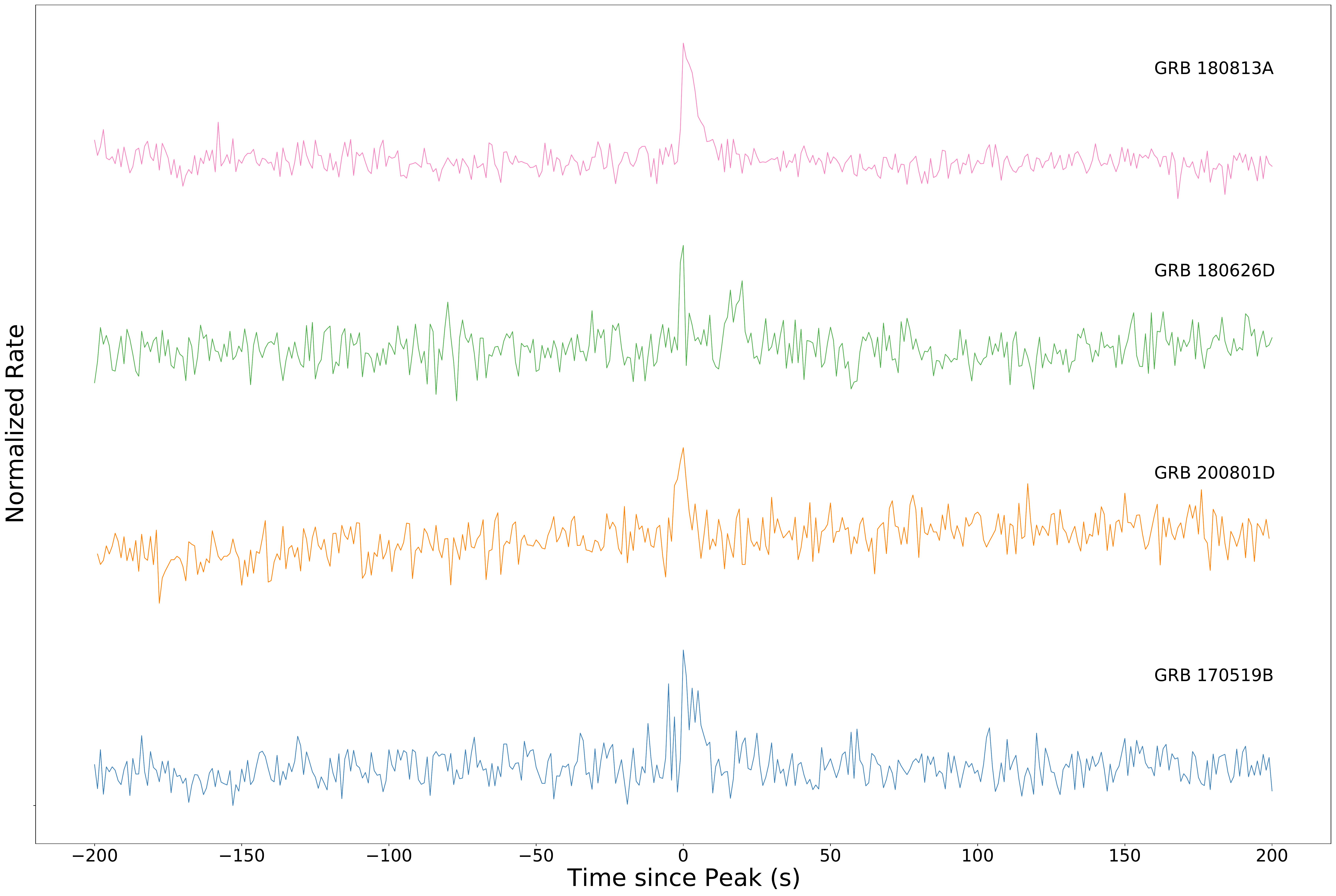}
    \caption{Veto lightcurves}\label{fig:discovery_lc_veto}
 \end{subfigure}
     \caption{The normalised lightcurves of GRBs discovered by CIFT that have not been reported by any instrument before (\S\ref{sec:new}). Each GRB light curve is normalised and labeled with the GRB name. Panel \subref{fig:unknown_lc_czti} shows normalised lightcurves for the GRBs detected in CZTI. The three sub-panels are with 0.1~s, 1~s and 10~s binning respectively, and each sub-panel is ordered by peak count rate above background, increasing from top to bottom. Panel \subref{fig:unknown_lc_veto} shows the normalised lightcurves of GRBs that were detected only in Veto. These are plotted with a 1~s binning, and are also ordered by peak count rate above background, increasing from top to bottom.}
%    \caption{\todo{Update y axis label to say ``normalised rate''} Normalised lightcurves of all the GRBs that CIFT discovered in 2017,2018, and 2019 with temporal binning increasing from left to right from 0.1s to 10s.}
    \label{fig:discoveredGRBlc}
\end{figure*}

\subsection{Known transients}\label{sec:pocmiss}
We detected 41 transients (referred as `Known') that had previously been reported by other instruments but had not been identified in CZTI or Veto data (Figure~\ref{fig:knownGRBlc}). These transients were matched to earlier reports in GCN Circulars\footnote{\url{https://gcn.gsfc.nasa.gov/gcn3_archive.html}}, {\em Fermi} GBM Burst Catalog\footnote{\url{https://heasarc.gsfc.nasa.gov/W3Browse/fermi/fermigbrst.html}} and the {\em Fermi} sub-threshold trigger lists\footnote{\url{https://gcn.gsfc.nasa.gov/fermi_gbm_subthresh_archive.html}}$^,$\footnote{\url{https://gammaray.nsstc.nasa.gov/gbm/science/sgrb_search.html}}.
Table \ref{tab:unknowntable} lists the key properties of these transients: a superevent ID, standard GRB name, trigger times (UTC), algorithms that detected the transient in CZTI or Veto data, temporal binning used in analysis, and the peak time (\asat\ time, measured as seconds since UT~2010-01-01 00:00:00). We then list the calculated parameters: the duration (\tnt), peak count rates above background, background count rates and total counts across all quadrants. We prefer using CZTI data to calculate these parameters. Even when our algorithms find a transient only in Veto detectors, we manually check if CZTI data can be used for calculation for uniformity. We use Veto data to calculate transient properties only if the transient is unseen in CZTI light curves. These cases are demarcated clearly in Table~\ref{tab:unknowntable}.

%The `algorithms' column has abbreviations `CS' for cumulative fraction cutoffs, `NS' for N-sigma and `TN' for Top-N algorithm listed after the data it was detected in (`CZTI:' and `Veto:'). The trigger times are specified in UTC and peak times in AstroSat time.

\subsection{CIFT discoveries}\label{sec:new}
We discovered 47 new transients that have not been reported by any instrument before. As in \S\ref{sec:pocmiss}, we show their light curves in Figure~\ref{fig:discoveredGRBlc}) and list properties in Table \ref{tab:discoverytable}. Six of these transients have been published already: GRB~180112B \citep{2018GCN.23511....1S}, GRB~190628B \citep{2019GCN.24972....1M}, GRB~191102A \citep{2019GCN.26378....1S}, GRB~191105B \citep{2019GCN.26376....1S}, GRB~191119A \citep{2019GCN.26268....1S}, and GRB~200817B \citep{2020GCN.28354....1S}.

\begin{figure}[!htp]
\centering
\begin{subfigure}{0.9\linewidth}
\centering
	\includegraphics[width=\columnwidth]{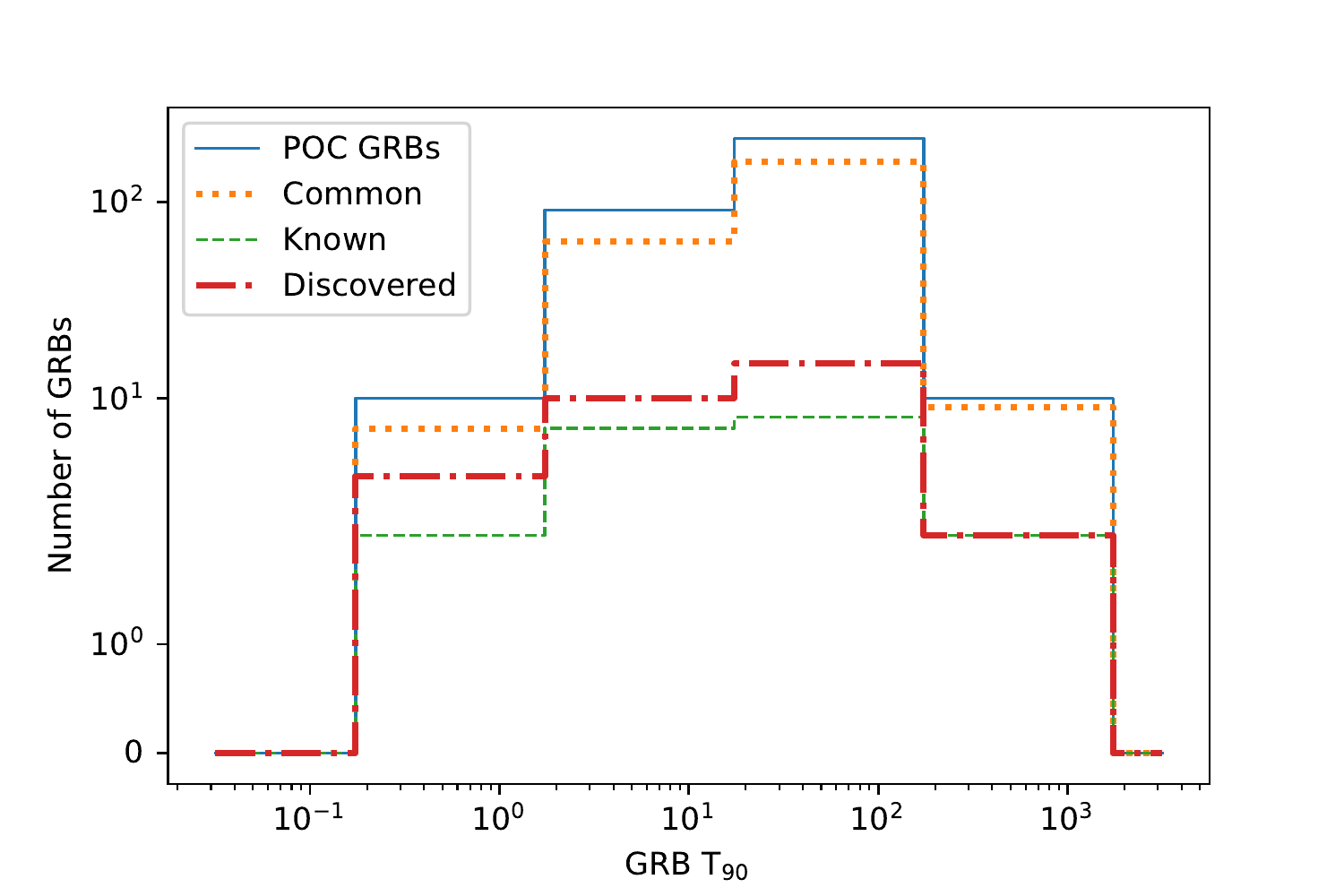}
	\caption{Distribution of \tnt\ values of all CZTI GRBs.}\label{fig:prop_t90}
\end{subfigure}

\begin{subfigure}{0.9\linewidth}
\centering
	\includegraphics[width=\columnwidth]{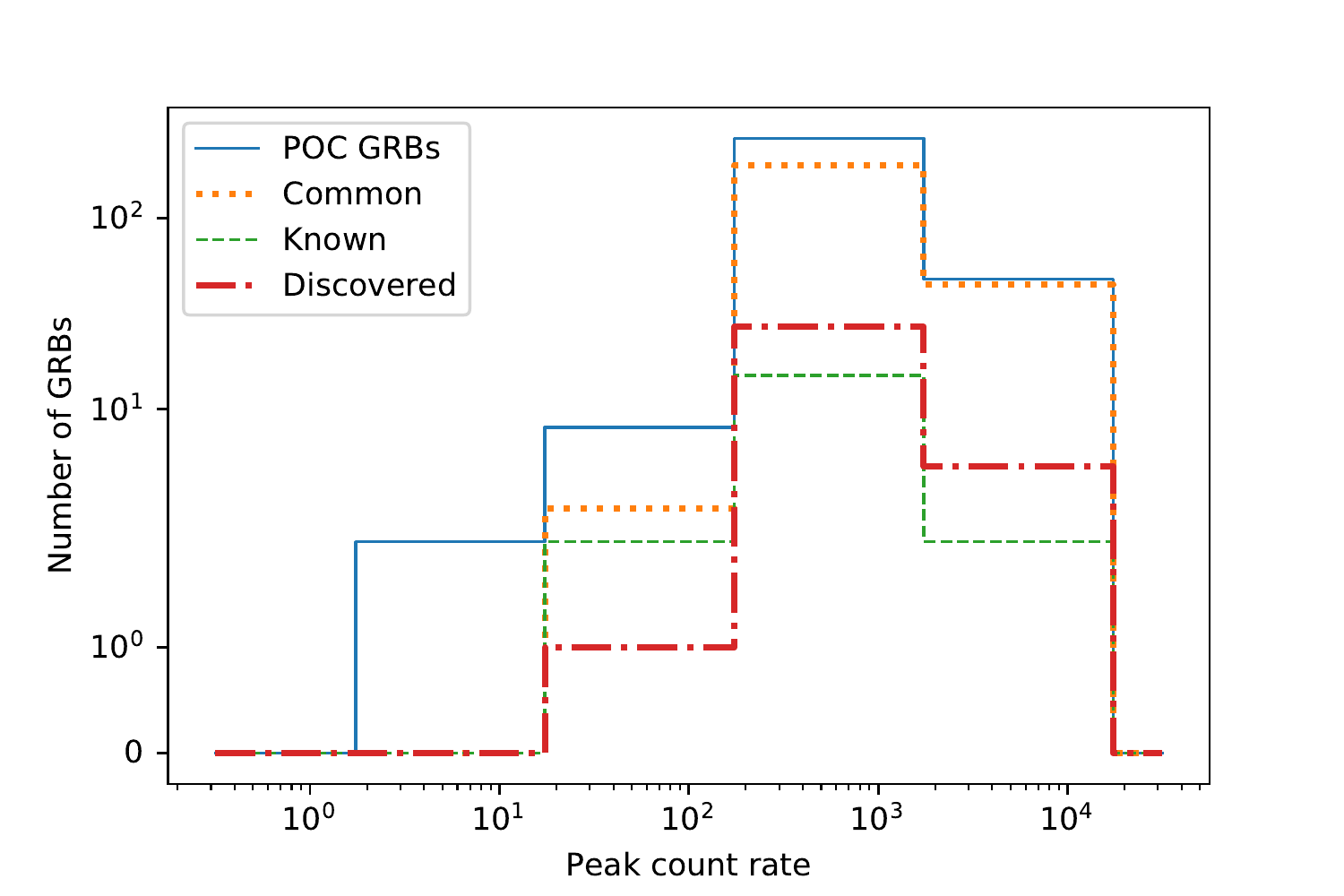}
	\caption{Distribution of peak count rates of transients in CZT detectors. Note that  high peak rates are often obtained for short duration transients analysed with 0.1~s binning.}\label{fig:prop_peak}
\end{subfigure}

\begin{subfigure}{0.9\linewidth}
\centering
	\includegraphics[width=\columnwidth]{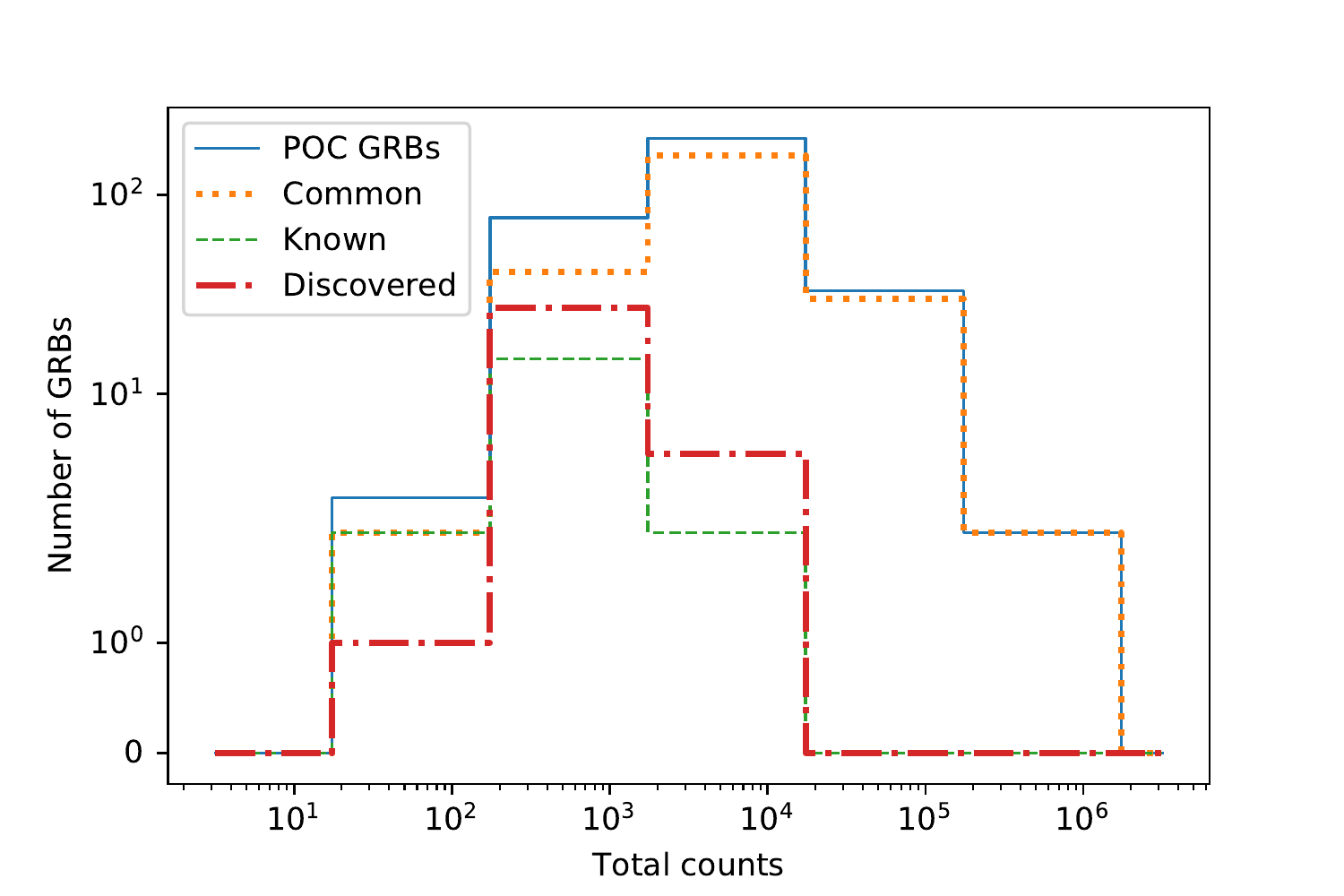}
	\caption{Distribution of total counts in CZT detectors.} \label{fig:prop_total}
\end{subfigure}
\caption{Comparing the distributions of the duration (\tnt), peak count rate, and total counts in three search classes. Blue lines (``POC GRBs'') denote transients detected in regular triggered searches and ML pipeline searches. Orange lines (``Common GRBs'') denote the transients CIFT detected among the ``POC GRBs''. Green lines (``Known GRBs'') denote transients that have been reported by other instruments (\S\ref{sec:pocmiss}) but missed by POC searches or ML pipeline, while red lines (``Discovered GRBs'') denote the new transients discovered with CIFT.}
\label{fig:properties}
\end{figure}

\subsection{Properties of new transients}\label{sec:properties}
The new transients detected by CIFT (\S\ref{sec:pocmiss} and \S\ref{sec:new}) span a wide range of properties.
%, but are not distinct from transients detected in in the past by the CZTI team in triggered and ML searches.
The shortest transient was GRB~200907A (\tnt\ = 0.13~s), while the longest was GRB~180809C with $T_{90}$ of 290~sec. GRB~200510B had the highest count rate above background (6461.6 count/s), while GRB~200906B had the lowest (53.7 count/s). Figure~\ref{fig:properties} shows the distributions of \tnt, peak count rate, and total counts for the four classes of transients: (a) those reported in the past by CZTI POC, (b) transients reported by POC which were also found by CIFT, (c) CIFT-detected transients reported by other instruments, and (d) new CIFT discoveries. We observe that all four classes have similar distributions of \tnt. A notable difference is seen in the total counts: transients with higher number of total CZT counts tend to be easily detected in regular triggered and ML searches. Also, GRBs with low peak count rates are more likely to be found in triggered searches undertaken by the POC but missed by CIFT. Note that although the three classes ``POC-GRBs'', ``Known GRBs'', and ``New Discovered'' are mutually exclusive, the distributions overlap well at the faint end of the distribution.

\edited{We find that about 10\% of all GRBs detected by CZTI are short GRBs, and the fraction remains the same for the 88 new bursts discovered with CIFT. The fraction of short GRBs is similar to the values for {\em Swift}-BAT \citep{2016ApJ...829....7L}, but smaller than the 26\% measured in {\em Fermi} \citep{2020ApJ...893...46V}. Here, we note an important caveat that we draw the line between short and long GRBs at the canonical value of $T_{90} = 2$~s, but it is known that this can be different for different instruments and will have to be measured separately for CZTI. As an illustration, if we adapt the {\em Fermi} boundary of 6.1~s, we find that about one-third of all CZTI GRBs are short GRBs.}

\subsection{Transients missed by CIFT}\label{sec:missedGRB}
\begin{figure}[hbtp]
    \centering
    \includegraphics[width=\columnwidth]{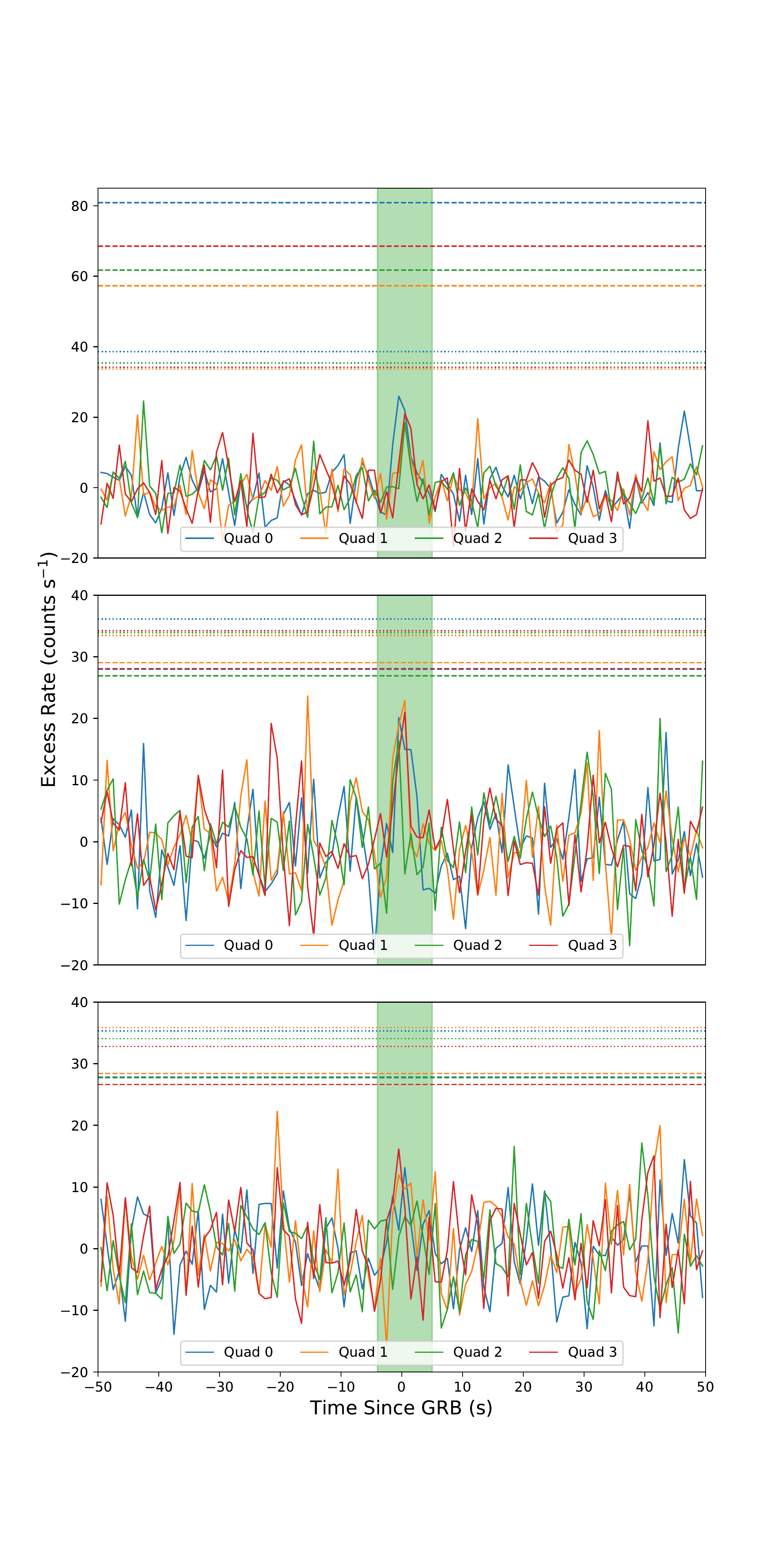}
    \caption{Diagnostic light curves for GRB~190605A. Top panel: 20--50~keV light curves for all four CZTI quadrants. The shaded green region denotes the GRB. Dashed and dotted lines denote the outlier threshold for CS and NS methods respectively, for each of the four quadrants. Middle panel: same as top panel, but for 50--100~keV. Bottom panel: same as top panel, but for 100--200~keV. The transient light curve looks similar in all quadrants, but it is too faint to qualify as an outlier in any of the methods.}
    \label{fig:diagnostic}
\end{figure}

Sixty-five GRBs that were found in regular triggered + ML searches were missed in the blind search with CIFT. Two of the missed GRBs were in \asat~slew orbits which were skipped while processing, as mentioned in \S\ref{sec:results}. We analysed the remaining cases to find the reasons why these were missed. The most common reason for the missed GRBs was that the transients were too faint in terms of their peak count rates. For instance, Figure~\ref{fig:diagnostic} shows the multi-quadrant, multi-band light curves for GRB~190605A. Visually, it is clear that the GRB is only weakly detected in all three search bands in CZTI data. In order to quantify this further, we calculated the count rates that would have been necessary to flag a data point as an outlier in the peak maps for this orbit. These rates for the CS method are shown with dashed lines, while the 5-$\sigma$ rates for NS are shown with dotted lines. It is clearly seen that the transient is well below these rates.

Such transients are rather easily confirmed by a human scanner inspecting the spectrogram and finding similar patterns in multiple quadrants. For quantitative analysis with say the CS method, the search window for a triggered search is usually set to 100~s, much smaller than the 4156~s window used in blind searches. This results in a lower cutoff rate, and will make more such fainter transients detectable in the current CIFT framework. Similarly, a smaller search window enables lowering the NS threshold from 5-$\sigma$ to 4-$\sigma$ or 3-$\sigma$ thanks to the fewer data points present, thereby increasing the odds of detecting fainter transients.

\begin{table*}[h]
\centering
\renewcommand\arraystretch{1.2}
\begin{tabular}{l| c c c c}
\toprule
\toprule
%   \rowcolor{gray!25}
  Algorithm & Candidates & \makecell{Common with triggered \\ or ML searches} & \makecell{Known transients} & New discoveries \\
  \midrule
%   \rowcolor{yellow!25}
  \multicolumn{1}{l}{\textbf{CZTI}} \\
%  \midrule
  Cutoff rate & 1290 & 206 & 16 & 30 \\
  NSigma & 2082 & 164 & 7 & 19\\
  TopN & 4199 & 210 & 19 & 30\\
  \midrule
%   \rowcolor{yellow!25}
  \multicolumn{1}{l}{\textbf{VETO}} \\
%  \midrule
  Cutoff rate & 10375 & 191 & 24 & 29\\
  NSigma & 10625 & 178 & 18 & 23\\
  TopN & 13993 & 222 & 34 & 32\\
  \midrule
%   \rowcolor{gray!25}
    \textbf{Total} & 22564 & 260 & 41 & 47 \\
  \bottomrule
\end{tabular}
%  \end{adjustbox}
   \caption{\label{tab:summary}Comparison of the three search algorithms running on  CZTI  and  VETO  for  different classifications  of the candidates identified. The `Candidates' column contains all potential transient candidates identified by our pipeline. The `Common with triggered or ML searches' column contains all GRBs that were originally detected by triggered or ML searches on CZTI data. The `Known transients' column contains all transients that had previously been reported by other instruments but had not been identified in CZTI or Veto data. The column `Discoveries' comprises of all transients that have not been reported by any instrument before. Common events in various methods are shown in Figure~\ref{fig:upset_CZTI}.}
\end{table*}
~

\section{Conclusions and future work}\label{sec:conc}
CZTI has proven itself to be a sensitive transient detector, but our searches had largely been limited to triggered searches. The ML pipeline \citep{Abraham2019} was the first major step towards detection of new transients with CZTI. The development of these algorithms, software, and the CIFT interface provide us with a powerful tool to extend our work further. Here, we have demonstrated the utility of this tool with the discovery of 88 new transients that had been missed by previous searches, including 47 transients that had not been detected by any mission to date. This brings the total CZTI tally to 413 GRBs in the first five years of its operation since launch, or about $\sim$ 83 per year. \edited{For comparison,  \textit{Swift} BAT detects $\sim $ 92 GRBs per year from on-board triggers \citep{2016ApJ...829....7L}, while \textit{Fermi} GBM detects $\sim$ 235 GRBs per year \citep{2020ApJ...893...46V}.}

%This brings the total CZTI tally to $\sim$83 GRBs per year, comparable to the $\sim$92 GRBs each year by {\em Swift} BAT \citep{2016ApJ...829....7L}. The sensitivity of CZTI to GRBs is discussed further in \citet{mate_this_issue}.

The CIFT framework is constantly evolving. It has been designed to make it easy to incorporate new features including search algorithms. We are working on metrics to quantify the statistical significance of a transient, so that we can lower the FPP. We have developed and tested a new search based on Bayesian Blocks \citep[BB; ][]{scargle2012}. We use the \texttt{astropy.stats.bayesian\textunderscore blocks} module to obtain block representations of de-trended light curves, and search for blocks that are 3-$\sigma$ outliers. These outliers then form the peak maps discussed in \S\ref{sec:bandsmaps}, so the BB search can easily be integrated into CIFT as a fourth algorithm. Preliminary testing has shown promising results with significantly lower false positive rates as compared to other algorithms. We will now run the BB search on the full data set.

Searches for fast transients also stand to benefit from other developments in CZTI data processing. New methods for rejecting noise from raw data \citep{ratheesh_this_issue} are improving the quality of light curves. These promise to lower the cut-off rates for CS by a factor of a few and will give a proportional boost to the count rate sensitivity of CZTI. Another notable change to be introduced is the non-removal of veto-tagged events. The default CZTI pipeline attributes coincident events between CZT and Veto detectors to charged particles, and discards them. In case of bright GRBs, large numbers of photons are incident both on CZT and Veto detectors, greatly increasing the chance coincidence rates. Since these are real GRB photons which should not be discarded, future CIFT-based searches will disable Veto-event filtering.

We have also added functionality to undertake specialised searches for X-ray counterparts to Fast Radio Bursts (FRBs) and Gravitational Wave (GW) sources. 
\edited{For instance, the magnetar source SGR 1935+2154 became active in early 2020, creating a series of bursts including one coincident with Fast Radio Burst \citep{2020arXiv200511071L}. We used the CIFT interface to efficiently search CZTI data for any bursts from this source. A first blind search was conducted with the default thresholds and we found three  bursts, coincident with times reported by other instruments. We then lowered the search thresholds and found an additional four bursts, corresponding to those reported by other missions. We are in the process of analysing properties of these CZTI-detected bursts, and the results will be reported separately (Raman et al., in prep).}   
We have also incorporated the ability to process GW localisation maps to calculate direction-dependent sensitivity. These features will streamline and boost the effort to search for X-ray counterparts to GW sources from the third observing run of advanced gravitational wave detectors \citep{abbott2020gwtc2}.

\section*{Acknowledgements}

CZT--Imager is built by a consortium of Institutes across India. The Tata Institute of Fundamental Research, Mumbai, led the effort with instrument design and development. Vikram Sarabhai Space Centre, Thiruvananthapuram provided the electronic design, assembly and testing. ISRO Satellite Centre (ISAC), Bengaluru provided the mechanical design, quality consultation and project management. The Inter University Centre for Astronomy and Astrophysics (IUCAA), Pune did the Coded Mask design, instrument calibration, and Payload Operation Centre. Space Application Centre (SAC) at Ahmedabad provided the analysis software. Physical Research Laboratory (PRL) Ahmedabad, provided the polarisation detection algorithm and ground calibration. A vast number of industries participated in the fabrication and the University sector pitched in by participating in the test and evaluation of the payload.

The Indian Space Research Organisation funded, managed and facilitated the project.

This work utilised various software including Python, AstroPy \citep{astropy}, NumPy \citep{numpy}, Matplotlib \citep{matplotlib}, \href{https://github.com/jnothman/upsetplot/}{UpSetPlot} \citep{upsetplot}, and \href{https://ngrok.com}{ngrok}.

\vspace{-1em}

\bibliography{biblio}

%%use \balance somewhere in the left column of the last page to balance the two columns in the end page

%%References section

%\begin{theunbibliography}{}
%\vspace{-1.5em}
%
%\bibitem{latexcompanion}
%Clark D. H., Caswell J. L. 1976, MNRAS, 174, 267
%\bibitem{latexcompanion}
%Dickey, J. M., Salpeter, E. E., Terzian, Y. 1978, Astrophys. J. Suppl. Ser., 36, 77
%\bibitem{latexcompanion}
%Radhakrishnan, G. C. {\em et al.} 1980, in Evans A., Bode M. F., eds, Non-Solar Gamma Rays (COSPAR), Pergamon Press, Oxford, p. 163
%\bibitem{latexcompanion}
%Starrfield S., Iliadis C., Hix W. R. 2008, in Bode M. F., Evans A., eds, Classical Novae, 2nd edition, Cambridge University Press, Cambridge, p. 77
%\bibitem{latexcompanion}
%Van Loon J. Th. 2008, in Evans A. et al., eds, R S Ophiuchi (2006) and the Recurrent Nova Phenomenon, ASP Conference Series, Volume 401, p. 90
%\bibitem{latexcompanion}
%Zwicky, F. 1957, Morphological Astronomy, Springer-Verlag, Berlin, p. 258
%
%
%
%\end{theunbibliography}

\onecolumn

\begingroup
\renewcommand\arraystretch{1.2}
\footnotesize
\begin{tabularx}{\linewidth}{l l l l l X r r r r}
	\caption{The table contains the calculated parameters for all `Known GRBs', which are the GRBs that had previously been reported by other instruments but had not been identified in CZTI or Veto data. The table is divided into three parts classified by what detector was used to detect (`Detected') and compute the parameters (`Analysed') given in the table. The column `SuperID' gives the name of the superevent identified by the pipeline. The column `GRB Name' contains the published name of the GRB, linked to the GRB report (more details in \ref{sec:pocmiss}). Several of these entries in the `GRB Name' column have a mark against their names, which gives the information on which quadrants were used for calculating the other parameters for that GRB. If there is no mark, then all four quadrants are used. Otherwise, marks `1', `2', `3', `4', and `5' refer to the quadrant sets -- `A,B,C', `A,B,D', `A,C,D', `B,C,D' and `C,D' respectively. The column `Algorithm' tells us what algorithms detected the GRB, where `TN', `NS', `CR' stands for the three algorithms -- TopN, N-sigma, and Cut-off rate respectively while `C' \& `V' are the two detectors -- CZTI and Veto. The time in \asat~seconds where the GRB was brightest is given in the column `Peak Time'. The bin size, that was used to generate the parameters -- T$_{90}$, Peak Count Rate above background, Background Rate and Total counts, is mentioned in the column `Bin'.} \\
	\toprule
	\toprule    
	{SuperID} & {GRB Name}& {Time} & {Algorithm}  & {\shortstack{Peak \\ Time}} & Bin & {T$_{90}$} & {\shortstack{Peak \\ Rate}} & {\shortstack{Bkg \\ Rate}}& {\shortstack{Total \\ Counts}}\\
% 	\midrule[0.001pt]
	{} & {}& {(UTC)} & {} & {(s)} & {(s)} & {(s)} & {(cps)} & {(cps)}& {(counts)}\\
	\midrule
	\endfirsthead
	\toprule
	{SuperID} & {GRB Name}& {Time} & {Algorithm} & {\shortstack{Peak \\ Time}} & Bin & {T$_{90}$} & {\shortstack{Peak \\ Rate}} & {\shortstack{Bkg \\ Rate}}& {\shortstack{Total \\ Counts}}\\
	{} & {}& {(UTC)} & {} & {(s)} & {(s)} & {(s)} & {(cps)} & {(cps)}& {(counts)}\\
	\midrule
	\endhead
	\midrule
	\multicolumn{8}{r}{\footnotesize(Contd.)}
	\endfoot
	\bottomrule
	\endlastfoot
	\input{unknown_table_latex.txt}
	\label{tab:unknowntable}
\end{tabularx}
\endgroup

\begingroup
\renewcommand\arraystretch{1.2}
\footnotesize
\begin{tabularx}{\textwidth}{l l l l l X r r r r}
	\caption{The table contains the calculated parameters for all `Discovered GRBs', which are all GRBs that have not been reported by any instrument before. The table is also divided into three parts classified by what detector was used to detect (`Detected') and compute the parameters (`Analysed') given in the table. The column `SuperID' gives the name of the superevent identified by the pipeline. Several of these entries in the `GRB Name' column have a mark against their names that tells what quadrants were used for calculating all other parameters for that GRB. If there is no mark, then all four quadrants are used. Otherwise, marks `1', `2', `3', `4', and `5' refer to the quadrant sets -- `A,B,C', `A,B,D', `A,C,D', `B,C,D' and `C,D' respectively. The column `Algorithm' tells us what algorithms detected the GRB, where `TN', `NS', `CR' stands for the three algorithms -- TopN, N-sigma, and Cut-off rate respectively whereas `C' \& `V' are the two detectors -- CZTI and Veto. The time in \asat~seconds where the GRB was brightest is given in the column `Peak Time'. The bin size, that was used to generate the parameters -- T$_{90}$, Peak Count Rate above background, Background Rate and Total counts, is mentioned in the column `Bin'.} \\
	\toprule
	\toprule
	{SuperID} & {GRB Name}& {Time} & {Algorithm} &  {\shortstack{Peak \\ Time}} & Bin & {T$_{90}$} & {\shortstack{Peak \\ Rate}} & {\shortstack{Bkg \\ Rate}}& {\shortstack{Total \\ Counts}}\\
	{} & {}& {(UTC)} & {} & {(s)} & {(s)} & {(s)} & {(cps)} & {(cps)}& {(counts)}\\
	%	{} & {}& {UTC} & {} & {(s)} & {} & {(s)} & {cps} & {cps}& {counts}\\
	\midrule
	\endfirsthead
	\toprule
	{SuperID} & {GRB Name}& {Time} & {Algorithm} & {\shortstack{Peak \\ Time}}& Bin  & {T$_{90}$} & {\shortstack{Peak \\ Rate}} & {\shortstack{Bkg \\ Rate}}& {\shortstack{Total \\ Counts}}\\
	{} & {}& {(UTC)} & {} & {(s)} & {(s)} & {(s)} & {(cps)} & {(cps)}& {(counts)}\\
	\midrule
	\endhead
	\midrule
	\multicolumn{8}{r}{\footnotesize(Contd.)}
	\endfoot
	\bottomrule
	\endlastfoot
	\input{discovery_table_latex.txt}
	\label{tab:discoverytable}
\end{tabularx}
\endgroup

% \begin{tabularx}{\linewidth}{c c r c l}
% \caption{Combined cut-offs for Cutoff rate and NSigma methods for each binning and band. These representative rates were calculated by using a month of data from each of the five years of data used in this study. Note that rates are in units of counts/sec, not counts per bin. %\label{tab:knownbutPOCmissed}
% \label{tab:combinedcut-off}}\\
% 	\toprule
% 	\toprule
% 	{Method}& {Binning} & {} & {Band-wise cutoff} & {}\\
% 	{} & {(s)} & 0 & 1 & 2\\
% 	\midrule
% 	\endfirsthead
% 	\bottomrule
% 	\endlastfoot
% % 	\input{cutoffs_czti.txt}
%     \textbf{CZTI} & & & &\\
% 	\midrule
% 	&0.1&1954&488&428\\
% 	{Cutoff rate}&{1.0}&263&107&102\\
% 	&{10.0}&28&22&22\\
% 	\midrule
% 	&{0.1}&396&410&407\\
% 	{NSigma}&{1.0}&137&133&131\\
% 	&10.0&41&38&38\\
	
% 	\toprule
% 	\toprule
% 	{Method}& {Binning} & {} & {Combined} & {}\\
% 	{} & {(s)} &  & cutoffs  & \\
% 	\midrule
% 	\textbf{VETO} & & & & \\
% 	\midrule
% 	\multirow{2}{*}{Cutoff rate}&1.0& & 319 & \\
% 	&10.0& &690 & \\
% 	\midrule
% 	\multirow{2}{*}{NSigma}&1.0&&394&\\
% 	&10.0&&1150&\\

% \end{tabularx}

\appendix

\twocolumn
\end{document}

%% file: unknown_table_latex.txt
\textbf{Detected:} & \textbf{CZTI} &  &  &  &  &  & & & \\
\textbf{Analyzed:} & \textbf{CZTI} &  &  &  &  &  & & & \\
\midrule
S196531116.0 & \href{https://gcn.gsfc.nasa.gov/gcn3/19248.gcn3}{GRB160324A\footnotemark[5]} & 15:58:34 & \makecell[l]{{\footnotesize C: CR, TN}\\ \footnotesize{V: CR, NS, TN}} & 196531117.5 & 1 & $48_{-17}^{+38}$ & $120_{-14}^{+30}$ & $209.2_{-1.0}^{+0.9}$ & $1889_{-589}^{+570}$\\
\midrule[0.01pt]
S200724370.0 & \href{https://heasarc.gsfc.nasa.gov/db-perl/W3Browse/w3hdprods.pl?files=P&Target=heasarc%5Ffermigbrst%7C%7C%7C%5F%5Frow%3D673%7C%7C&Coordinates=Equatorial&Equinox=2000}{GRB160512A\footnotemark[3]} & 04:46:08 & \makecell[l]{{\footnotesize C: CR, TN}\\ \footnotesize{V: CR}} & 200724382.5 & 1 & $26_{-4}^{+6}$ & $237_{-26}^{+43}$ & $336_{-3}^{+2}$ & $1854_{-382}^{+347}$\\
\midrule[0.01pt]
S204095177.0 & \href{https://gcn.gsfc.nasa.gov/gcn/gcn3/19594.gcn3}{GRB160620A} & 05:06:15 & \makecell[l]{{\footnotesize C: CR, NS, TN}\\ \footnotesize{V: CR, NS, TN}} & 204095178.4 & 0.1 & $0.99_{-0.17}^{+0.62}$ & $1385_{-222}^{+218}$ & $447_{-10}^{+8}$ & $714_{-103}^{+101}$\\
\midrule[0.01pt]
S219728856.0 & \href{https://gcn.gsfc.nasa.gov/gcn3/20283.gcn3}{GRB161218A} & 3:47:34 & \makecell[l]{{\footnotesize C: CR, NS, TN}\\ \footnotesize{V: NS, TN}} & 219728856.1 & 0.1 & $6_{-1}^{+1}$ & $457_{-18}^{+190}$ & $423_{-12}^{+10}$ & $1134_{-276}^{+227}$\\
\midrule[0.01pt]
S224556541.0 & \href{https://heasarc.gsfc.nasa.gov/db-perl/W3Browse/w3hdprods.pl?files=P&Target=heasarc%5Ffermigbrst%7C%7C%7C%5F%5Frow%3D1418%7C%7C&Coordinates=Equatorial&Equinox=2000}{GRB170212A\footnotemark[2]} & 00:48:59 & \makecell[l]{{\footnotesize C: CR, NS, TN}\\ \footnotesize{V: CR, TN}} & 224556540.2 & 0.1 & $4_{-1}^{+2}$ & $353_{-39}^{+150}$ & $316_{-6}^{+8}$ & $536_{-189}^{+135}$\\
\midrule[0.01pt]
S229730480.0 & \href{https://heasarc.gsfc.nasa.gov/db-perl/W3Browse/w3hdprods.pl?files=P&Target=heasarc%5Ffermigbrst%7C%7C%7C%5F%5Frow%3D2035%7C%7C&Coordinates=Equatorial&Equinox=2000}{GRB170412B} & 22:01:18 & \makecell[l]{{\footnotesize C: CR, NS, TN}\\ \footnotesize{V: None}} & 229730479.5 & 1 & $51_{-17}^{+7}$ & $318_{-47}^{+44}$ & $425_{-3}^{+3}$ & $2589_{-689}^{+667}$\\
\midrule[0.01pt]
S230047201.0 & \href{https://heasarc.gsfc.nasa.gov/db-perl/W3Browse/w3hdprods.pl?files=P&Target=heasarc%5Ffermigbrst%7C%7C%7C%5F%5Frow%3D276%7C%7C&Coordinates=Equatorial&Equinox=2000}{GRB170416A\footnotemark[4]} & 13:59:59 & \makecell[l]{{\footnotesize C: TN}\\ \footnotesize{V: CR, NS, TN}} & 230047201.2 & 0.1 & $5.9_{-1.4}^{+0.2}$ & $288_{-26}^{+141}$ & $333_{-7}^{+8}$ & $551_{-135}^{+115}$\\
\midrule[0.01pt]
S239852052.0 & \href{https://heasarc.gsfc.nasa.gov/db-perl/W3Browse/w3hdprods.pl?files=P&Target=heasarc%5Ffermigbrst%7C%7C%7C%5F%5Frow%3D2747%7C%7C&Coordinates=Equatorial&Equinox=2000}{GRB170808C} & 01:34:10 & \makecell[l]{{\footnotesize C: CR, NS, TN}\\ \footnotesize{V: None}} & 239852052.4 & 0.1 & $4.5_{-0.5}^{+0.4}$ & $772_{-86}^{+196}$ & $422_{-12}^{+8}$ & $1106_{-179}^{+185}$\\
\midrule[0.01pt]
S250719370.0 & \href{https://heasarc.gsfc.nasa.gov/db-perl/W3Browse/w3hdprods.pl?files=P&Target=heasarc%5Ffermigbrst%7C%7C%7C%5F%5Frow%3D1489%7C%7C&Coordinates=Equatorial&Equinox=2000}{GRB171211B} & 20:16:08 & \makecell[l]{{\footnotesize C: CR, TN}\\ \footnotesize{V: TN}} & 250719471.5 & 1 & $135_{-8}^{+8}$ & $278_{-37}^{+41}$ & $358_{-2}^{+2}$ & $4827_{-641}^{+697}$\\
\midrule[0.01pt]
S254815405.0 & \href{https://heasarc.gsfc.nasa.gov/db-perl/W3Browse/w3hdprods.pl?files=P&Target=heasarc%5Ffermigbrst%7C%7C%7C%5F%5Frow%3D1226%7C%7C&Coordinates=Equatorial&Equinox=2000}{GRB180128B\footnotemark[1]} & 06:03:23 & \makecell[l]{{\footnotesize C: CR, TN}\\ \footnotesize{V: NS, TN}} & 254815407.2 & 0.1 & $5.7_{-0.8}^{+0.7}$ & $297_{-17}^{+195}$ & $466_{-13}^{+9}$ & $814_{-200}^{+190}$\\
\midrule[0.01pt]
S269585348.0 & \href{https://heasarc.gsfc.nasa.gov/db-perl/W3Browse/w3hdprods.pl?files=P&Target=heasarc%5Ffermigbrst%7C%7C%7C%5F%5Frow%3D449%7C%7C&Coordinates=Equatorial&Equinox=2000}{GRB180718C} & 04:49:06 & \makecell[l]{{\footnotesize C: CR, TN}\\ \footnotesize{V: None}} & 269585341.5 & 1 & $10_{-4}^{+6}$ & $217_{-40}^{+38}$ & $363_{-2}^{+3}$ & $1024_{-230}^{+157}$\\
\midrule[0.01pt]
S271117239.0 & \href{https://heasarc.gsfc.nasa.gov/db-perl/W3Browse/w3hdprods.pl?files=P&Target=heasarc%5Ffermigbrst%7C%7C%7C%5F%5Frow%3D107%7C%7C&Coordinates=Equatorial&Equinox=2000}{GRB180804A} & 22:20:37 & \makecell[l]{{\footnotesize C: CR, NS, TN}\\ \footnotesize{V: CR, NS, TN}} & 271117238.8 & 0.1 & $5.2_{-0.8}^{+0.6}$ & $695_{-43}^{+209}$ & $434_{-8}^{+7}$ & $1313_{-142}^{+152}$\\
\midrule[0.01pt]
S272640643.0 & \href{https://heasarc.gsfc.nasa.gov/db-perl/W3Browse/w3hdprods.pl?files=P&Target=heasarc%5Ffermigbrst%7C%7C%7C%5F%5Frow%3D211%7C%7C&Coordinates=Equatorial&Equinox=2000}{GRB180822B\footnotemark[2]} & 13:30:41 & \makecell[l]{{\footnotesize C: CR, TN}\\ \footnotesize{V: None}} & 272640638.0 & 10 & $126_{-110}^{+64}$ & $72_{-10}^{+11}$ & $352.4_{-0.7}^{+0.4}$ & $1589_{-705}^{+712}$\\
\midrule[0.01pt]
S272942427.0 & \href{https://heasarc.gsfc.nasa.gov/db-perl/W3Browse/w3hdprods.pl?files=P&Target=heasarc%5Ffermigbrst%7C%7C%7C%5F%5Frow%3D904%7C%7C&Coordinates=Equatorial&Equinox=2000}{GRB180826A\footnotemark[2]} & 01:20:25 & \makecell[l]{{\footnotesize C: CR, TN}\\ \footnotesize{V: CR, NS, TN}} & 272942423.0 & 10 & $130_{-14}^{+11}$ & $119_{-13}^{+12}$ & $493.5_{-1.1}^{+0.8}$ & $6535_{-551}^{+565}$\\
\midrule[0.01pt]
S279646944.0 & \href{https://heasarc.gsfc.nasa.gov/db-perl/W3Browse/w3hdprods.pl?files=P&Target=heasarc%5Ffermigbrst%7C%7C%7C%5F%5Frow%3D1739%7C%7C&Coordinates=Equatorial&Equinox=2000}{GRB181111A} & 15:42:22 & \makecell[l]{{\footnotesize C: TN}\\ \footnotesize{V: None}} & 279646935.5 & 1 & $14_{-3}^{+3}$ & $192_{-42}^{+41}$ & $480_{-3}^{+3}$ & $1138_{-210}^{+176}$\\
\midrule[0.01pt]
S280064235.0 & \href{https://heasarc.gsfc.nasa.gov/db-perl/W3Browse/w3hdprods.pl?files=P&Target=heasarc%5Ffermigbrst%7C%7C%7C%5F%5Frow%3D880%7C%7C&Coordinates=Equatorial&Equinox=2000}{GRB181116A} & 11:37:13 & \makecell[l]{{\footnotesize C: CR, TN}\\ \footnotesize{V: CR, NS, TN}} & 280064267.5 & 1 & $74_{-4}^{+9}$ & $316_{-35}^{+46}$ & $454_{-2}^{+2}$ & $5805_{-626}^{+590}$\\
\midrule[0.01pt]
S280166126.0 & \href{https://heasarc.gsfc.nasa.gov/db-perl/W3Browse/w3hdprods.pl?files=P&Target=heasarc%5Ffermigbrst%7C%7C%7C%5F%5Frow%3D2018%7C%7C&Coordinates=Equatorial&Equinox=2000}{GRB181117A\footnotemark[3]} & 15:55:24 & \makecell[l]{{\footnotesize C: CR, TN}\\ \footnotesize{V: CR, NS, TN}} & 280166128.5 & 1 & $12_{-3}^{+3}$ & $133_{-23}^{+37}$ & $328_{-3}^{+2}$ & $921_{-174}^{+208}$\\
\midrule[0.01pt]
S326222331.0 & \href{https://gcn.gsfc.nasa.gov/gcn/gcn3/27682.gcn3}{GRB200503B} & 17:18:50 & \makecell[l]{{\footnotesize C: TN}\\ \footnotesize{V: None}} & 326222323.5 & 1 & $62_{-12}^{+11}$ & $155_{-25}^{+43}$ & $155_{-25}^{+43}$ & $2372_{-696}^{+656}$\\
\midrule[0.01pt]
S337184665.0 & \href{https://gcn.gsfc.nasa.gov/gcn3/28378.gcn3}{GRB200907A\footnotemark[3]} & 14:24:24 & \makecell[l]{{\footnotesize C: CR, NS, TN}\\ \footnotesize{V: CR, NS, TN}} & 337184664.7 & .01 & $0.13_{-0.02}^{+0.01}$ & $4718_{-1079}^{+1176}$ & $476_{-47}^{+16}$ & $260_{-29}^{+33}$\\
\midrule[0.01pt]
\midrule
\textbf{Detected:} & \textbf{Veto} &  &  &  &  &  & & & \\
\textbf{Analyzed:} & \textbf{CZTI} &  &  &  &  &  & & & \\
\midrule
S185452630.0 & \href{https://heasarc.gsfc.nasa.gov/db-perl/W3Browse/w3hdprods.pl?files=P&Target=heasarc%5Ffermigbrst%7C%7C%7C%5F%5Frow%3D172%7C%7C&Coordinates=Equatorial&Equinox=2000}{GRB151117A} & 10:37:08 & \makecell[l]{{\footnotesize C: None}\\ \footnotesize{V: CR, TN}} & 185452622.8 & 0.1 & $5_{-1}^{+2}$ & $267_{-9}^{+168}$ & $383_{-8}^{+5}$ & $493_{-155}^{+150}$\\
\midrule[0.01pt]
S190192926.0 & \href{https://gcn.gsfc.nasa.gov/gcn3/18864.gcn3}{GRB160111A} & 07:22:04 & \makecell[l]{{\footnotesize C: None}\\ \footnotesize{V: TN}} & 190192925.5 & .01 & $_{}^{}$ & $903_{-170}^{+186}$ & $431_{-10}^{+8}$ & $126_{-88}^{+84}$\\
\midrule[0.01pt]
S222932540.0 & \href{https://heasarc.gsfc.nasa.gov/db-perl/W3Browse/w3hdprods.pl?files=P&Target=heasarc%5Ffermigbrst%7C%7C%7C%5F%5Frow%3D2830%7C%7C&Coordinates=Equatorial&Equinox=2000}{GRB170124A\footnotemark[4]} & 05:42:18 & \makecell[l]{{\footnotesize C: None}\\ \footnotesize{V: CR, TN}} & 222932533.6 & 0.1 & $10.3_{-1.2}^{+0.5}$ & $264_{-17}^{+160}$ & $328_{-6}^{+6}$ & $897_{-154}^{+187}$\\
\midrule[0.01pt]
S241696600.0 & \href{https://heasarc.gsfc.nasa.gov/db-perl/W3Browse/w3hdprods.pl?files=P&Target=heasarc%5Ffermigbrst%7C%7C%7C%5F%5Frow%3D767%7C%7C&Coordinates=Equatorial&Equinox=2000}{GRB170829B\footnotemark[2]} & 09:56:38 & \makecell[l]{{\footnotesize C: None}\\ \footnotesize{V: TN}} & 241696628.5 & 1 & $45_{-3}^{+4}$ & $182_{-36}^{+35}$ & $352_{-2}^{+3}$ & $2320_{-430}^{+297}$\\
\midrule[0.01pt]
S250224944.0 & \href{https://heasarc.gsfc.nasa.gov/db-perl/W3Browse/w3hdprods.pl?files=P&Target=heasarc%5Ffermigbrst%7C%7C%7C%5F%5Frow%3D25%7C%7C&Coordinates=Equatorial&Equinox=2000}{GRB171206A\footnotemark[1]} & 2:55:42 & \makecell[l]{{\footnotesize C: None}\\ \footnotesize{V: TN}} & 250224944.7 & 0.1 & $3_{-1}^{+1}$ & $272_{-13}^{+170}$ & $477_{-11}^{+9}$ & $297_{-155}^{+156}$\\
\midrule[0.01pt]
S254611360.0 & \href{https://heasarc.gsfc.nasa.gov/db-perl/W3Browse/w3hdprods.pl?files=P&Target=heasarc%5Ffermigbrst%7C%7C%7C%5F%5Frow%3D345%7C%7C&Coordinates=Equatorial&Equinox=2000}{GRB180125A\footnotemark[3]} & 21:22:38 & \makecell[l]{{\footnotesize C: None}\\ \footnotesize{V: TN}} & 254611351.5 & 1 & $24_{-6}^{+14}$ & $95.5_{-0.6}^{+51.8}$ & $447_{-3}^{+3}$ & $1352_{-394}^{+385}$\\
\midrule[0.01pt]
S257333099.0 & \href{https://gcn.gsfc.nasa.gov/notices_gbm_sub/541329904.fermi}{GRB180226A} & 09:24:57 & \makecell[l]{{\footnotesize C: None}\\ \footnotesize{V: NS, TN}} & 257333099.3 & 0.1 & $0.78_{-0.53}^{+0.57}$ & $346_{-50}^{+159}$ & $340_{-7}^{+8}$ & $152_{-57}^{+61}$\\
\midrule[0.01pt]
S264915479.0 & \href{https://heasarc.gsfc.nasa.gov/db-perl/W3Browse/w3hdprods.pl?files=P&Target=heasarc%5Ffermigbrst%7C%7C%7C%5F%5Frow%3D1918%7C%7C&Coordinates=Equatorial&Equinox=2000}{GRB180525A} & 03:37:57 & \makecell[l]{{\footnotesize C: None}\\ \footnotesize{V: TN}} & 264915479.1 & .01 & $0.16_{-0.05}^{+0.02}$ & $1128_{-120}^{+798}$ & $487_{-41}^{+20}$ & $88_{-25}^{+27}$\\
\midrule[0.01pt]
S266353140.0 & \href{https://heasarc.gsfc.nasa.gov/db-perl/W3Browse/w3hdprods.pl?files=P&Target=heasarc%5Ffermigbrst%7C%7C%7C%5F%5Frow%3D2255%7C%7C&Coordinates=Equatorial&Equinox=2000}{GRB180610C} & 18:58:58 & \makecell[l]{{\footnotesize C: None}\\ \footnotesize{V: CR, TN}} & 266353130.5 & 1 & $18_{-6}^{+1}$ & $122_{-6}^{+47}$ & $391_{-2}^{+3}$ & $1164_{-185}^{+180}$\\
\midrule[0.01pt]
S267371600.0 & \href{https://heasarc.gsfc.nasa.gov/db-perl/W3Browse/w3hdprods.pl?files=P&Target=heasarc%5Ffermigbrst%7C%7C%7C%5F%5Frow%3D2885%7C%7C&Coordinates=Equatorial&Equinox=2000}{GRB180622B\footnotemark[2]} & 13:53:18 & \makecell[l]{{\footnotesize C: None}\\ \footnotesize{V: TN}} & 267371591.5 & 1 & $27_{-5}^{+5}$ & $242_{-38}^{+50}$ & $707_{-4}^{+3}$ & $2498_{-474}^{+457}$\\
\midrule[0.01pt]
S273953547.0 & \href{https://heasarc.gsfc.nasa.gov/db-perl/W3Browse/w3hdprods.pl?files=P&Target=heasarc%5Ffermigbrst%7C%7C%7C%5F%5Frow%3D117%7C%7C&Coordinates=Equatorial&Equinox=2000}{GRB180906B} & 18:12:25 & \makecell[l]{{\footnotesize C: None}\\ \footnotesize{V: CR, TN}} & 273953548.1 & 0.1 & $5.8_{-1.9}^{+0.5}$ & $250_{-35}^{+208}$ & $517_{-12}^{+10}$ & $569_{-177}^{+183}$\\
\midrule[0.01pt]
S288536007.0 & \href{https://heasarc.gsfc.nasa.gov/db-perl/W3Browse/w3hdprods.pl?files=P&Target=heasarc%5Ffermigbrst%7C%7C%7C%5F%5Frow%3D2721%7C%7C&Coordinates=Equatorial&Equinox=2000}{GRB190222B} & 12:53:25 & \makecell[l]{{\footnotesize C: None}\\ \footnotesize{V: CR, NS, TN}} & 288536007.5 & 1 & $10.4_{-2.9}^{+0.6}$ & $214_{-24}^{+40}$ & $341_{-3}^{+2}$ & $1030_{-145}^{+136}$\\
\midrule[0.01pt]
\midrule
\textbf{Detected:} & \textbf{Veto} &  &  &  &  &  & & & \\
\textbf{Analyzed:} & \textbf{Veto} &  &  &  &  &  & & & \\
\midrule
S197170443.0 & \href{https://heasarc.gsfc.nasa.gov/db-perl/W3Browse/w3hdprods.pl?files=P&Target=heasarc%5Ffermigbrst%7C%7C%7C%5F%5Frow%3D751%7C%7C&Coordinates=Equatorial&Equinox=2000}{GRB160401A} & 01:34:01 & \makecell[l]{{\footnotesize C: None}\\ \footnotesize{V: CR, NS, TN}} & 197170442.4 & 1 & $14_{-4}^{+14}$ & $385_{-59}^{+60}$ & $952_{-4}^{+5}$ & $1626_{-581}^{+459}$\\
\midrule[0.01pt]
S207052803.0 & \href{https://heasarc.gsfc.nasa.gov/db-perl/W3Browse/w3hdprods.pl?files=P&Target=heasarc%5Ffermigbrst%7C%7C%7C%5F%5Frow%3D1829%7C%7C&Coordinates=Equatorial&Equinox=2000}{GRB160724A} & 10:40:03 & \makecell[l]{{\footnotesize C: None}\\ \footnotesize{V: CR, NS, TN}} & 207052810.8 & 1 & $16_{-7}^{+9}$ & $980_{-86}^{+80}$ & $1560_{-6}^{+5}$ & $3098_{-666}^{+632}$\\
\midrule[0.01pt]
S224077190.0 & \href{https://gcn.gsfc.nasa.gov/gcn3/20639.gcn3}{GRB170206C} & 11:39:48 & \makecell[l]{{\footnotesize C: None}\\ \footnotesize{V: CR, NS, TN}} & 224077218.4 & 1 & $_{}^{}$ & $218_{-30}^{+68}$ & $1511_{-6}^{+5}$ & $2193_{-638}^{+594}$\\
\midrule[0.01pt]
S249867183.0 & \href{https://gcn.gsfc.nasa.gov/notices_gbm_sub/533863531.fermi}{GRB171201} & 23:33:01 & \makecell[l]{{\footnotesize C: None}\\ \footnotesize{V: TN}} & 249867182.8 & 1 & $_{}^{}$ & $798_{-86}^{+76}$ & $1555_{-7}^{+5}$ & $853_{-185}^{+166}$\\
\midrule[0.01pt]
S258277398.0 & \href{https://heasarc.gsfc.nasa.gov/db-perl/W3Browse/w3hdprods.pl?files=P&Target=heasarc%5Ffermigbrst%7C%7C%7C%5F%5Frow%3D2260%7C%7C&Coordinates=Equatorial&Equinox=2000}{GRB180309A\footnotemark[1]} & 07:43:16 & \makecell[l]{{\footnotesize C: None}\\ \footnotesize{V: CR, TN}} & 258277397.7 & 1 & $25_{-8}^{+5}$ & $310_{-61}^{+60}$ & $1041_{-4}^{+3}$ & $2484_{-416}^{+389}$\\
\midrule[0.01pt]
S264108750.0 & \href{https://heasarc.gsfc.nasa.gov/db-perl/W3Browse/w3hdprods.pl?files=P&Target=heasarc%5Ffermigbrst%7C%7C%7C%5F%5Frow%3D2861%7C%7C&Coordinates=Equatorial&Equinox=2000}{GRB180515A} & 19:32:28 & \makecell[l]{{\footnotesize C: None}\\ \footnotesize{V: CR, NS, TN}} & 264108761.5 & 1 & $20_{-2}^{+4}$ & $441_{-50}^{+71}$ & $1555_{-6}^{+5}$ & $6091_{-582}^{+528}$\\
\midrule[0.01pt]
S267468139.0 & \href{https://gcn.gsfc.nasa.gov/gcn/gcn3/22828.gcn3}{GRB180623A} & 16:42:17 & \makecell[l]{{\footnotesize C: None}\\ \footnotesize{V: CR, NS, TN}} & 267468140.9 & 1 & $64_{-4}^{+2}$ & $518_{-48}^{+70}$ & $1163_{-4}^{+4}$ & $6986_{-1088}^{+773}$\\
\midrule[0.01pt]
S278544286.0 & \href{https://heasarc.gsfc.nasa.gov/db-perl/W3Browse/w3hdprods.pl?files=P&Target=heasarc%5Ffermigbrst%7C%7C%7C%5F%5Frow%3D835%7C%7C&Coordinates=Equatorial&Equinox=2000}{GRB181029A} & 21:24:44 & \makecell[l]{{\footnotesize C: None}\\ \footnotesize{V: CR, NS, TN}} & 278544285.6 & 1 & $10_{-5}^{+1}$ & $392_{-70}^{+66}$ & $1309_{-5}^{+5}$ & $1286_{-307}^{+278}$\\
\midrule[0.01pt]
S280573750.0 & \href{https://heasarc.gsfc.nasa.gov/db-perl/W3Browse/w3hdprods.pl?files=P&Target=heasarc%5Ffermigbrst%7C%7C%7C%5F%5Frow%3D2257%7C%7C&Coordinates=Equatorial&Equinox=2000}{GRB181122A\footnotemark[2]} & 09:09:08 & \makecell[l]{{\footnotesize C: None}\\ \footnotesize{V: CR, TN}} & 280573745.5 & 1 & $44_{-13}^{+3}$ & $237_{-45}^{+60}$ & $1113_{-5}^{+4}$ & $2634_{-623}^{+566}$\\
\midrule[0.01pt]
S289607820.0 & \href{https://heasarc.gsfc.nasa.gov/db-perl/W3Browse/w3hdprods.pl?files=P&Target=heasarc%5Ffermigbrst%7C%7C%7C%5F%5Frow%3D1450%7C%7C&Coordinates=Equatorial&Equinox=2000}{GRB190306B} & 22:36:58 & \makecell[l]{{\footnotesize C: None}\\ \footnotesize{V: CR, TN}} & 289607814.3 & 1 & $_{}^{}$ & $199_{-2}^{+89}$ & $1563_{-7}^{+6}$ & $3597_{-1499}^{+1454}$\\
\midrule[0.01pt]

%% file: discovery_table_latex.txt
\textbf{Detected:} & \textbf{CZTI} &  &  &  &  &  & & & \\
\textbf{Analyzed:} & \textbf{CZTI} &  &  &  &  &  & & & \\
\midrule
S184303433.0 & GRB151104A & 3:23:51 & \makecell[l]{{\footnotesize C: CR, NS, TN}\\ \footnotesize{V: CR, NS, TN}} & 184303438.5 & 1 & $68_{-4}^{+2}$ & $911_{-43}^{+62}$ & $474_{-4}^{+3}$ & $28049_{-827}^{+744}$\\
\midrule[0.01pt]
S184512194.0 & GRB151106A\footnotemark[3] & 13:23:12 & \makecell[l]{{\footnotesize C: CR, NS, TN}\\ \footnotesize{V: None}} & 184512190.5 & 1 & $39.4_{-25.5}^{+0.8}$ & $203_{-25}^{+39}$ & $319_{-3}^{+2}$ & $1596_{-292}^{+285}$\\
\midrule[0.01pt]
S195263948.0 & GRB160309A\footnotemark[5] & 23:59:06 & \makecell[l]{{\footnotesize C: CR, TN}\\ \footnotesize{V: None}} & 195263943.0 & 10 & $67_{-16}^{+27}$ & $62_{-9}^{+8}$ & $211.3_{-0.6}^{+0.4}$ & $2034_{-618}^{+521}$\\
\midrule[0.01pt]
S197184964.0 & GRB160401C\footnotemark[1] & 05:36:02 & \makecell[l]{{\footnotesize C: CR, NS, TN}\\ \footnotesize{V: CR, NS, TN}} & 197184965.2 & 0.1 & $3.1_{-1.0}^{+0.9}$ & $588_{-20}^{+189}$ & $386_{-10}^{+10}$ & $1090_{-180}^{+152}$\\
\midrule[0.01pt]
S199212123.0 & GRB160424B & 16:42:01 & \makecell[l]{{\footnotesize C: CR, NS, TN}\\ \footnotesize{V: None}} & 199212122.5 & 1 & $43.6_{-2.2}^{+0.7}$ & $691_{-62}^{+54}$ & $554_{-3}^{+4}$ & $4093_{-511}^{+342}$\\
\midrule[0.01pt]
S199449121.0 & GRB160427A & 10:31:59 & \makecell[l]{{\footnotesize C: CR, TN}\\ \footnotesize{V: CR, NS, TN}} & 199449124.5 & 1 & $34_{-12}^{+3}$ & $375_{-50}^{+46}$ & $476_{-3}^{+4}$ & $2669_{-372}^{+317}$\\
\midrule[0.01pt]
S203963626.0 & GRB160618A & 16:33:44 & \makecell[l]{{\footnotesize C: CR, NS, TN}\\ \footnotesize{V: CR, NS, TN}} & 203963626.6 & .01 & $0.27_{-0.01}^{+0.01}$ & $4918_{-366}^{+1328}$ & $444_{-48}^{+17}$ & $877_{-61}^{+61}$\\
\midrule[0.01pt]
S215358438.0 & GRB161028A & 13:47:16 & \makecell[l]{{\footnotesize C: CR, TN}\\ \footnotesize{V: None}} & 215358429.5 & 1 & $25.7_{-0.3}^{+0.6}$ & $284_{-36}^{+44}$ & $434_{-4}^{+3}$ & $1883_{-271}^{+267}$\\
\midrule[0.01pt]
S218385140.0 & GRB161202C\footnotemark[4] & 14:32:27 & \makecell[l]{{\footnotesize C: CR, TN}\\ \footnotesize{V: CR, NS, TN}} & 218385206.5 & 1 & $146_{-4}^{+1}$ & $131_{-20}^{+35}$ & $326_{-2}^{+2}$ & $3346_{-848}^{+767}$\\
\midrule[0.01pt]
S226302000.0 & GRB170304B & 05:39:58 & \makecell[l]{{\footnotesize C: CR, NS, TN}\\ \footnotesize{V: CR, NS, TN}} & 226302047.5 & 1 & $41_{-3}^{+2}$ & $5091_{-126}^{+120}$ & $486_{-2}^{+2}$ & $23464_{-511}^{+495}$\\
\midrule[0.01pt]
S232990228.0 & GRB170520B & 15:30:26 & \makecell[l]{{\footnotesize C: CR}\\ \footnotesize{V: CR, NS, TN}} & 232990228.1 & 0.1 & $2.3_{-0.5}^{+0.8}$ & $506_{-61}^{+176}$ & $490_{-10}^{+11}$ & $551_{-162}^{+101}$\\
\midrule[0.01pt]
S239432831.0 & GRB170803F & 05:07:09 & \makecell[l]{{\footnotesize C: CR, NS, TN}\\ \footnotesize{V: CR, NS, TN}} & 239432833.4 & 0.1 & $6.0_{-0.4}^{+0.3}$ & $344_{-68}^{+172}$ & $452_{-13}^{+8}$ & $1494_{-227}^{+209}$\\
\midrule[0.01pt]
S242255598.0 & GRB170904B & 21:13:16 & \makecell[l]{{\footnotesize C: CR, NS, TN}\\ \footnotesize{V: NS, TN}} & 242255598.6 & 0.1 & $3.4_{-0.5}^{+0.4}$ & $966_{-175}^{+208}$ & $553_{-12}^{+10}$ & $1424_{-160}^{+199}$\\
\midrule[0.01pt]
S247846852.0 & GRB171108C\footnotemark[3] & 14:20:50 & \makecell[l]{{\footnotesize C: CR, NS, TN}\\ \footnotesize{V: None}} & 247846851.7 & 0.1 & $0.88_{-0.26}^{+0.48}$ & $915_{-143}^{+183}$ & $374_{-9}^{+6}$ & $326_{-63}^{+69}$\\
\midrule[0.01pt]
S253483949.0 & GRB180112B & 20:12:27 & \makecell[l]{{\footnotesize C: TN}\\ \footnotesize{V: CR, TN}} & 253483948.3 & 0.1 & $9_{-3}^{+2}$ & $335_{-3}^{+178}$ & $409_{-6}^{+5}$ & $794_{-159}^{+210}$\\
\midrule[0.01pt]
S257929386.0 & GRB180305B & 07:03:04 & \makecell[l]{{\footnotesize C: CR, NS, TN}\\ \footnotesize{V: CR, NS, TN}} & 257929428.5 & 1 & $32_{-5}^{+2}$ & $769_{-59}^{+56}$ & $436_{-3}^{+2}$ & $5170_{-315}^{+374}$\\
\midrule[0.01pt]
S264583245.0 & GRB180521B & 07:20:43 & \makecell[l]{{\footnotesize C: CR, TN}\\ \footnotesize{V: None}} & 264583239.5 & 1 & $15_{-9}^{+4}$ & $167_{-24}^{+46}$ & $529_{-3}^{+3}$ & $1026_{-289}^{+287}$\\
\midrule[0.01pt]
S271507716.0 & GRB180809C & 10:48:34 & \makecell[l]{{\footnotesize C: CR, NS, TN}\\ \footnotesize{V: CR, TN}} & 271507720.0 & 10 & $290_{-52}^{+22}$ & $132_{-14}^{+13}$ & $535_{-2}^{+1}$ & $8438_{-1464}^{+1314}$\\
\midrule[0.01pt]
S286741010.0 & GRB190201A\footnotemark[2] & 18:16:48 & \makecell[l]{{\footnotesize C: CR, TN}\\ \footnotesize{V: CR, TN}} & 286741004.5 & 1 & $8.0_{-2.9}^{+0.4}$ & $136_{-34}^{+38}$ & $369_{-4}^{+2}$ & $558_{-119}^{+122}$\\
\midrule[0.01pt]
S288677154.0 & GRB190224A & 04:05:52 & \makecell[l]{{\footnotesize C: CR, NS, TN}\\ \footnotesize{V: None}} & 288677164.5 & 1 & $12.9_{-0.4}^{+0.4}$ & $843_{-67}^{+63}$ & $606_{-5}^{+4}$ & $5397_{-261}^{+254}$\\
\midrule[0.01pt]
S298744222.0 & GRB190620B & 16:30:20 & \makecell[l]{{\footnotesize C: CR, NS, TN}\\ \footnotesize{V: None}} & 298744221.9 & 0.1 & $4_{-1}^{+2}$ & $824_{-55}^{+238}$ & $549_{-15}^{+12}$ & $1504_{-426}^{+362}$\\
\midrule[0.01pt]
S299391822.0 & GRB190628B\footnotemark[3] & 04:23:40 & \makecell[l]{{\footnotesize C: CR, TN}\\ \footnotesize{V: None}} & 299391815.5 & 1.0 & $32_{-24}^{+15}$ & $196_{-42}^{+40}$ & $405_{-3}^{+3}$ & $1177_{-513}^{+410}$\\
\midrule[0.01pt]
S310393113.0 & GRB191102B\footnotemark[1] & 12:18:31 & \makecell[l]{{\footnotesize C: CR, TN}\\ \footnotesize{V: None}} & 310393104.5 & 0.1 & $4.2_{-1.2}^{+0.8}$ & $279_{-25}^{+189}$ & $392_{-10}^{+7}$ & $583_{-124}^{+126}$\\
\midrule[0.01pt]
S310393113.0 & GRB191102A & 12:18:31 & \makecell[l]{{\footnotesize C: CR, TN}\\ \footnotesize{V: None}} & 310393105.0 & 0.1 & $5_{-3}^{+9}$ & $307_{-30}^{+208}$ & $513_{-11}^{+6}$ & $731_{-420}^{+313}$\\
\midrule[0.01pt]
S310630593.0 & GRB191105B & 06:16:31 & \makecell[l]{{\footnotesize C: CR, NS, TN}\\ \footnotesize{V: CR, NS, TN}} & 310630601.8 & 0.1 & $13.2_{-0.6}^{+0.4}$ & $665_{-77}^{+202}$ & $739_{-12}^{+7}$ & $2200_{-343}^{+340}$\\
\midrule[0.01pt]
S311856067.0 & GRB191119A & 10:41:05 & \makecell[l]{{\footnotesize C: CR, NS, TN}\\ \footnotesize{V: CR, NS, TN}} & 311856067.0 & .01 & $0.15_{-0.02}^{+0.03}$ & $2702_{-230}^{+1130}$ & $449_{-42}^{+18}$ & $270_{-34}^{+39}$\\
\midrule[0.01pt]
S317162956.0 & GRB200119B & 20:49:13 & \makecell[l]{{\footnotesize C: CR, TN}\\ \footnotesize{V: CR, NS, TN}} & 317162955.5 & 1 & $_{}^{}$ & $195_{-22}^{+52}$ & $524_{-4}^{+4}$ & $1593_{-1522}^{+1232}$\\
\midrule[0.01pt]
S324009902.0 & GRB200408B & 2:44:59 & \makecell[l]{{\footnotesize C: CR, NS, TN}\\ \footnotesize{V: None}} & 324009901.5 & 1 & $_{}^{}$ & $217_{-43}^{+45}$ & $494_{-4}^{+4}$ & $590_{-531}^{+731}$\\
\midrule[0.01pt]
S326787080.0 & GRB200510B & 6:11:17 & \makecell[l]{{\footnotesize C: CR, NS, TN}\\ \footnotesize{V: None}} & 326787121.8 & .01 & $0.29_{-0.01}^{+0.01}$ & $8318_{-366}^{+1896}$ & $520_{-50}^{+20}$ & $1695_{-75}^{+84}$\\
\midrule[0.01pt]
S329736162.0 & GRB200613C & 09:22:40 & \makecell[l]{{\footnotesize C: CR, TN}\\ \footnotesize{V: CR, NS, TN}} & 329736159.5 & 1 & $11_{-1}^{+2}$ & $132_{-3}^{+58}$ & $618_{-3}^{+4}$ & $1190_{-236}^{+208}$\\
\midrule[0.01pt]
S329806842.0 & GRB200613B\footnotemark[2] & 09:22:40 & \makecell[l]{{\footnotesize C: CR, TN}\\ \footnotesize{V: CR, NS, TN}} & 329806843.5 & 1 & $24_{-10}^{+7}$ & $104_{-8}^{+44}$ & $433_{-3}^{+3}$ & $1257_{-306}^{+288}$\\
\midrule[0.01pt]
S334929280.0 & GRB200812A & 11:54:37 & \makecell[l]{{\footnotesize C: CR, NS, TN}\\ \footnotesize{V: CR, NS, TN}} & 334929324.5 & 1 & $10_{-5}^{+2}$ & $1677_{-81}^{+72}$ & $449_{-3}^{+5}$ & $4415_{-224}^{+194}$\\
\midrule[0.01pt]
S335340170.0 & GRB200817B & 06:02:48 & \makecell[l]{{\footnotesize C: CR, NS, TN}\\ \footnotesize{V: CR, NS, TN}} & 335340230.5 & 1.0 & $23_{-5}^{+10}$ & $455_{-43}^{+48}$ & $490_{-2}^{+3}$ & $3847_{-507}^{+378}$\\
\midrule[0.01pt]
\midrule
\textbf{Detected:} & \textbf{Veto} &  &  &  &  &  & & & \\
\textbf{Analyzed:} & \textbf{CZTI} &  &  &  &  &  & & & \\
\midrule
S213125894.0 & GRB161002A & 17:38:12 & \makecell[l]{{\footnotesize C: None}\\ \footnotesize{V: CR, TN}} & 213125894.5 & .01 & $0.87_{-0.31}^{+0.14}$ & $991_{-56}^{+792}$ & $413_{-41}^{+34}$ & $235_{-105}^{+116}$\\
\midrule[0.01pt]
S228861170.0 & GRB170402C\footnotemark[1] & 20:32:48 & \makecell[l]{{\footnotesize C: None}\\ \footnotesize{V: CR, TN}} & 228861062.0 & 10 & $119_{-41}^{+68}$ & $82_{-9}^{+11}$ & $360.9_{-0.9}^{+0.8}$ & $5072_{-1232}^{+1150}$\\
\midrule[0.01pt]
S242743550.0 & GRB170910B & 12:45:48 & \makecell[l]{{\footnotesize C: None}\\ \footnotesize{V: TN}} & 242743547.1 & 0.1 & $2.6_{-1.8}^{+0.2}$ & $400_{-94}^{+164}$ & $471_{-10}^{+9}$ & $259_{-106}^{+107}$\\
\midrule[0.01pt]
S255979768.0 & GRB180210C\footnotemark[3] & 17:29:26 & \makecell[l]{{\footnotesize C: None}\\ \footnotesize{V: CR, NS, TN}} & 255979759.5 & 1 & $7_{-1}^{+3}$ & $43_{-6}^{+45}$ & $350_{-3}^{+3}$ & $110_{-511}^{+462}$\\
\midrule[0.01pt]
S267689714.0 & GRB180626D & 06:15:12 & \makecell[l]{{\footnotesize C: None}\\ \footnotesize{V: CR}} & 267689735.5 & 1 & $23_{-3}^{+6}$ & $133_{-10}^{+51}$ & $529_{-3}^{+3}$ & $1082_{-344}^{+314}$\\
\midrule[0.01pt]
S284139623.0 & GRB190102B\footnotemark[1] & 15:40:21 & \makecell[l]{{\footnotesize C: None}\\ \footnotesize{V: TN}} & 284139622.4 & 0.1 & $3.1_{-0.6}^{+0.5}$ & $252_{-12}^{+154}$ & $345_{-10}^{+7}$ & $309_{-86}^{+88}$\\
\midrule[0.01pt]
S285215230.0 & GRB190115A\footnotemark[2] & 02:27:08 & \makecell[l]{{\footnotesize C: None}\\ \footnotesize{V: CR, TN}} & 285215220.5 & 1 & $19_{-2}^{+1}$ & $162_{-24}^{+40}$ & $384_{-3}^{+2}$ & $1225_{-194}^{+187}$\\
\midrule[0.01pt]
S329776004.0 & GRB200613D\footnotemark[3] & 20:26:42 & \makecell[l]{{\footnotesize C: None}\\ \footnotesize{V: NS}} & 329776008.5 & 1 & $23_{-10}^{+4}$ & $97_{-17}^{+35}$ & $334_{-3}^{+3}$ & $925_{-247}^{+224}$\\
\midrule[0.01pt]
S329801810.0 & GRB200614B\footnotemark[4] & 03:36:47 & \makecell[l]{{\footnotesize C: None}\\ \footnotesize{V: NS, TN}} & 329801807.5 & 1 & $38.3_{-24.2}^{+0.5}$ & $112_{-6}^{+47}$ & $396_{-3}^{+2}$ & $1306_{-294}^{+379}$\\
\midrule[0.01pt]
S337054619.0 & GRB200906B & 02:16:56 & \makecell[l]{{\footnotesize C: None}\\ \footnotesize{V: CR, NS}} & 337054499.5 & 1 & $35_{-9}^{+14}$ & $110_{-6}^{+45}$ & $462_{-2}^{+2}$ & $2033_{-564}^{+690}$\\
\midrule[0.01pt]
S337615721.0 & GRB200912A & 14:08:38 & \makecell[l]{{\footnotesize C: None}\\ \footnotesize{V: CR, NS, TN}} & 337615720.8 & 0.1 & $2.6_{-1.3}^{+0.0}$ & $415_{-56}^{+169}$ & $497_{-10}^{+9}$ & $396_{-112}^{+83}$\\
\midrule[0.01pt]
\midrule
\textbf{Detected:} & \textbf{Veto} &  &  &  &  &  & & & \\
\textbf{Analyzed:} & \textbf{Veto} &  &  &  &  &  & & & \\
\midrule
S232918513.0 & GRB170519B & 19:35:11 & \makecell[l]{{\footnotesize C: None}\\ \footnotesize{V: CR, TN}} & 232918512.2 & 1 & $14_{-4}^{+6}$ & $377_{-64}^{+71}$ & $1583_{-7}^{+6}$ & $2251_{-448}^{+400}$\\
\midrule[0.01pt]
S271825254.0 & GRB180813A & 03:00:52 & \makecell[l]{{\footnotesize C: None}\\ \footnotesize{V: CR, NS, TN}} & 271825253.3 & 1 & $16_{-7}^{+9}$ & $506_{-60}^{+72}$ & $1464_{-6}^{+5}$ & $3144_{-493}^{+434}$\\
\midrule[0.01pt]
S333985240.0 & GRB200801D & 13:40:38 & \makecell[l]{{\footnotesize C: None}\\ \footnotesize{V: CR, TN}} & 333985233.0 & 1 & $6_{-3}^{+1}$ & $283_{-51}^{+72}$ & $1687_{-6}^{+6}$ & $1198_{-272}^{+251}$\\
\midrule[0.01pt]